\documentclass[usenatbib]{mnras}
\usepackage{amssymb}
\usepackage{dcolumn}
\usepackage{bm}
\usepackage{graphicx}
\usepackage{aas_macros}
\usepackage[caption=false]{subfig}
\usepackage{amsmath}
\usepackage{mathtools}

\begin{document}
\label{firstpage}
\title[GAMA and SAMI's Little Blue Spheroids]{Star-Forming, Rotating Spheroidal Galaxies in the GAMA and SAMI Surveys}

\author[Moffett et al.]{Amanda J. Moffett,$^{1,2}$$^\dagger$ Steven Phillipps,$^3$ Aaron S. G. Robotham,$^4$ Simon P. Driver,$^4$
\newauthor Malcolm N. Bremer,$^3$ Luca Cortese,$^{4,5}$ O. Ivy Wong,$^4$ Sarah Brough,$^6$ Michael J. I.
\newauthor Brown,$^7$ Julia J. Bryant,$^{5,8,9}$ Christopher J. Conselice,$^{10}$ Scott M. Croom,$^{5,6}$ Koshy
\newauthor George,$^{11,22}$ Greg Goldstein,$^{12}$ Michael Goodwin,$^9$ Benne W. Holwerda,$^{13}$ Andrew
\newauthor M. Hopkins,$^{15}$ Iraklis S. Konstantopoulos,$^{14}$ Jon S. Lawrence,$^{15}$ Nuria P. F. Lorente,$^{9}$
\newauthor Anne M. Medling,$^{16,17,18}$ Matt S. Owers,$^{12}$ Kevin A. Pimbblet,$^{19}$ Samuel N. Richards,$^{20}$
\newauthor Sarah M. Sweet,$^{5,21}$ and Jesse van de Sande$^{5,8}$ \\
$^1$Department of Physics and Astronomy, Vanderbilt University, PMB \#401807 2401 Vanderbilt Place, Nashville TN 37240, USA\\
$^2$Department of Physics and Astronomy, University of North Georgia, 3820 Mundy Mill Rd., Oakwood GA 30566, USA\\
$^3$School of Physics, University of Bristol, Tyndall Avenue, Bristol BS8 1TL, UK\\
$^4$ICRAR, The University of Western Australia, 35 Stirling Highway, Crawley WA 6009, Australia\\
$^5$ARC Centre of Excellence for All Sky Astrophysics in 3 Dimensions (ASTRO 3D)\\
$^6$School of Physics, University of New South Wales, NSW 2052, Australia\\
$^7$School of Physics and Astronomy, Monash University, VIC 3800, Australia\\
$^8$Sydney Institute for Astronomy, School of Physics, A28, The University of Sydney, NSW, 2006, Australia \\
$^{9}$Australian Astronomical Optics, 105 Delhi Rd, North Ryde, NSW 2113, Australia\\
$^{10}$School of Physics \& Astronomy, University of Nottingham, Nottingham NG7 2RD, UK\\
$^{11}$Indian Institute of Astrophysics, 2nd Block, Koramangala, Bangalore - 560034, India\\
$^{12}$Department of Physics and Astronomy, Macquarie University, NSW 2109, Australia\\
$^{13}$Department of Physics and Astronomy, 102 Natural Science Building, University of Louisville, Louisville KY 40292, USA\\
$^{14}$Atlassian, 341 George St Sydney, NSW 2000, Australia\\
$^{15}$Australian Astronomical Optics, Macquarie University, 105 Delhi Rd, North Ryde, NSW 2113, Australia\\
$^{16}$Ritter Astrophysical Research Center University of Toledo Toledo, OH 43606, USA\\
$^{17}$Research School for Astronomy \& Astrophysics Australian National University Canberra, ACT 2611, Australia\\
$^{18}$Hubble Fellow \\
$^{19}$E. A. Milne Centre for Astrophysics, University of Hull, Cottingham Road, Kingston-upon-Hull, HU6 7RX, UK\\
$^{20}$SOFIA Science Center, USRA, NASA Ames Research Center, Building N232, M/S 232-12, P.O. Box 1, Moffett Field, CA 94035-0001, USA\\
$^{21}$Centre for Astrophysics and Supercomputing, Swinburne University of Technology, PO Box 218, Hawthorn, VIC 3122, Australia\\
$^{22}$Department of Physics, Christ University, Hosur Road, Bangalore 560029, India\\ 
$^\dagger${E-mail: amanda.moffett@ung.edu} }

\maketitle

\begin{abstract}
  The Galaxy And Mass Assembly (GAMA) survey has morphologically identified a class of ``Little Blue Spheroid'' (LBS) galaxies whose relationship to other classes of galaxies we now examine in detail. Considering a sample of 868 LBSs, we find that such galaxies display similar but not identical colours, specific star formation rates, stellar population ages, mass-to-light ratios, and metallicities to Sd-Irr galaxies. We also find that LBSs typically occupy environments of even lower density than those of Sd-Irr galaxies, where $\sim$65\% of LBS galaxies live in isolation. Using deep, high-resolution imaging from VST KiDS and the new Bayesian, two-dimensional galaxy profile modeling code PROFIT, we further examine the detailed structure of LBSs and find that their S\'ersic indices, sizes, and axial ratios are compatible with those of low-mass elliptical galaxies. We then examine SAMI Galaxy survey integral field emission line kinematics for a subset of 62 LBSs and find that the \emph{majority} (42) of these galaxies display ordered rotation with the remainder displaying disturbed/non-ordered dynamics. Finally, we consider potential evolutionary scenarios for a population with this unusual combination of properties, concluding that LBSs are likely formed by a mixture of merger and accretion processes still recently active in low-redshift dwarf populations. We also infer that if LBS-like galaxies were subjected to quenching in a rich environment, they would plausibly resemble cluster dwarf ellipticals.
\end{abstract}


\section{Introduction}

In the last three decades, increasingly large samples of galaxies have been surveyed in terms of their photometric and spectroscopic properties (e.g. 2dFGRS, \citealp{2df}; SDSS, \citealp{SDSS}; GAMA, \citealp{GAMA}). Such surveys have provided the means to quantify the characteristic properties of different classes of galaxies, and a primary method of dividing galaxies into classes with similar characteristics is through morphological classification. The interest in galaxy morphology is motivated in part by the apparent link between a galaxy's structure and its likely formation history, with spheroidal structures generally thought to result from dissipationless processes such as dry mergers (e.g., \citealp{Cole2000}) and disk-like structures thought to result from dissipational gas physics processes (e.g., \citealp{FE1980}). While it is likely that the dynamics of a galaxy provide a more direct probe of their formation histories, morphological classification has one advantage in its feasibility for significantly larger populations of galaxies.

The large-scale morphological classification of survey samples include work carried out by survey teams themselves (e.g., \citealp{Kelvin_class}, \citealp{vismorph}), by ``citizen scientists,'' as in the Galaxy Zoo project (e.g., \citealp{GZ}, \citealp{GZ2}), and through automated classification schemes such as CAS (\citealp{CAS}), Gini-M20 \citep{Lotz04}, and deep learning algorithms \citep{HC15}. Such work has presented an opportunity to identify new or rare types of galaxies, for instance the SDSS discoveries of ``Green Peas'' \citep{peas} and isolated compact ellipticals \citep{HPP13}.

Another recently identified class of galaxy is the blue but morphologically early-type galaxy, which includes the blue ellipticals of \citet{blueE} and the blue E/S0s of \citet{KGB} and \citet{Schaw09}. Here we discuss another such class of galaxies known as the ``little blue spheroids'' identified in the Galaxy And Mass Assembly (GAMA) survey. Note that each of these classes are likely to differ in detailed properties due to the differing limits and selections of their origin samples. In particular, the mass ranges of these blue early-type populations differ signficantly, where the \citet{Schaw09} population are approximately $L^{*}$ galaxies, the \citet{KGB} population reaches into the dwarf mass regime with stellar mass $\sim$10$^{8}$~{\rm M$_{\odot}$}, and GAMA's ``little blue spheroids'' extend down to stellar mass $\sim$10$^{7}$~{\rm M$_{\odot}$}.

These blue early-type galaxies may also overlap with the well-studied, nearby blue compact dwarf (BCD) galaxy population \citep{BCD}, which although identified in a number of ways by different authors, share the characteristics of being blue (typically judged via optical colour of the core), compact (judged by high $B$-band surface brightness of the core), and dwarf (low mass but often traced by an optical luminosity cut; \citealp{BCDdef}). These requirements can select for galaxies with morphological similarity to the aforementioned blue early types, however observed BCDs come in a variety of shapes, from those that appear as a purely spheroidal core to those with both smooth and clumpy outer envelopes (see e.g., \citealp{BCD_morph}; \citealp{BCDdef}).

As part of the GAMA survey, visual morphological classifications were completed for a sample of galaxies out to z$=$0.06, by utilising three-colour ($giH$) postage-stamp images. Among these objects, \citet{Kelvin_class} identified a class that they called ``little blue spheroids'' (LBS), which appeared to lie outside the expected range of morphological types. This class of galaxies was defined by the multiple observer visual impression of compact and round morphology along with blue colour (also judged visually) and comprised 7.4\% of the classified sample. \citet{vismorph} subsequently expanded this visual morphology classification using the same visual classifiers of \citet{Kelvin_class} and approximately doubled the sample size by extending to a larger GAMA phase two sample with a fainter magnitude limit (r $<$ 19.8 compared to r $<$ 19.4 mag) and a larger sky area (180 deg$^2$ compared to 144 deg$^2$). LBS galaxies made up 11.5\% of this expanded sample, which contained a larger proportion of faint/lower mass objects. \citet{bdmfunc} also estimated that these galaxies make up around 1\% of the total low redshift stellar mass in GAMA. Note that since this classification is based purely on galaxy images as they are observed with a variety of on-sky projection angles, LBS galaxies may or may not actually represent spheroids in three dimensions. We treat the structural and dynamical similarity of this LBS class to spheroidal galaxies as a matter of investigation in this work.

The closest relatives to these LBS galaxies may be ``normal'' dwarf elliptical, or dE, galaxies, which share their apparent spheroidal appearance but typically not their blue colour. Dwarf elliptical galaxies are distinguished from their giant elliptical counterparts not only by virtue of lower mass (or fainter magnitude) but also by a profile shape that is typically more shallow, closer to an exponential profile than the steep de Vaucouleurs profile characteristic of giant ellipticals (e.g., \citealp{dE_prof}; \citealp{dE_prof2}; \citealp{dE_class}; \citealp{dE_review}). Much like a giant elliptical, a typical dE galaxy appears red in optical three-colour images similar to those used for classification in the GAMA survey (see e.g., \citealp{morph_review} for examples), and most dEs are characterized by old stellar population ages with little evidence for recent star formation. Most known dEs are also found in relatively rich environments, and dEs have been found to be the most numerous type of galaxy in several galaxy clusters (see review of \citealp{dE_review} and references therein).

In the present paper we explore the detailed properties of the galaxies classified as LBSs. We compare these to the properties of other types of low-mass galaxies in order to investigate the current and past relationship between LBSs and other classes of low-mass galaxies. We first introduce our sample and data in \S \ref{data}. We then consider the basic properties of LBS galaxies as derived from various GAMA-based catalogues in \S \ref{prop}, finding a strong similarity in stellar population properties and environments between GAMA LBSs and Sd-Irrs. We next analyze the detailed structural properties of LBSs using deep, high resolution optical imaging from the VST KiDS survey (\citealp{KiDS}) in \S \ref{struc}. We find that LBSs have quantitatively similar structure to low-mass ellipticals as might be expected from their qualitatively judged morphology. We then examine the emission line kinematics of a subsample of LBSs observed by the SAMI Galaxy survey \citep{SAMI_ins} in \S \ref{SAMIcomp}, finding that contrary to the expectation from their spheroid-like structure, most LBSs are at least marginally rotation-dominated galaxies. Finally in \S \ref{conc} we summarize our results and discuss their implications for the likely origins and future evolution of LBS galaxies, concluding that LBSs likely emerge from a mixture of galaxy-galaxy interaction and accretion processes and could form a plausible progenitor population for dwarf ellipticals.

Throughout this work we use a standard concordance cosmology, i.e. $H_0 = 70$~km~s$^{-1}$Mpc$^{-1}$, $\Omega_m = 0.3$, $\Omega_{\Lambda} = 0.7$, as in other GAMA data products.

\begin{figure*}
\begin{tabular}{ccc}
\subfloat[]{\includegraphics[width = 2in]{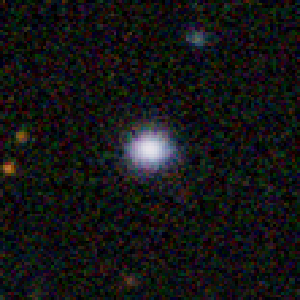}} &
\subfloat[]{\includegraphics[width = 2in]{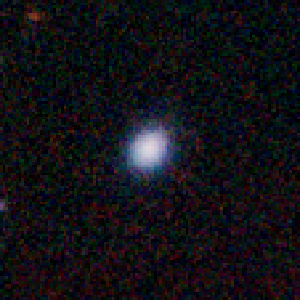}} &
\subfloat[]{\includegraphics[width = 2in]{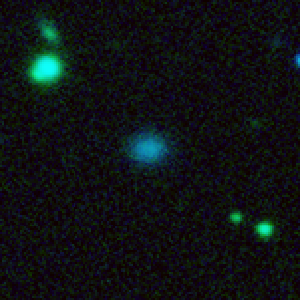}} 
\end{tabular}
\caption{Representative LBS galaxies (GAMA IDs G300372, G418795, and G417568) in their original three-colour ($giH$) classification images. Classification images are 30kpc on a side in size \citep{vismorph}, and all images are scaled using the same algorithm (tanh scaling) such that scaling differences reflect the changing dynamic range in each image (for example, the scaling in the far right panel is affected by a nearby bright point source).}
\label{fig:classims}
\end{figure*}

\section{GAMA Sample and Data Products}
\label{data}

The GAMA (Galaxy And Mass Assembly) survey is a combined spectroscopic and multi-wavelength imaging survey of five patches of the sky, with total area 286 square degrees, down to a magnitude limit of $r=19.8$, with spectroscopic observations from the AAOmega spectrograph on the Anglo-Australian Telescope (see \citealp{GAMAdr2} for a recent summary of GAMA Data Release 2). The three GAMA equatorial regions, amounting to a total sky area of 180 deg$^2$, were selected for this study due to the availability of visual morphology classifications and spectroscopic redshift survey completeness $>98$\% to $r < 19.8$ mag \citep{GAMAdr2}. The survey was based originally on catalogued SDSS photometry, but this has since been reprocessed and homogenized to give improved magnitudes (\citealp{GAMAphot1}; \citealp{lambdar}). In addition to the the basic photometric data and the survey's spectroscopic data \citep{GAMAspec}, GAMA catalogues also provide a wide range of derived properties (as described by \citealp{GAMAdr2}), such as stellar population and dust extinction parameters, inferred masses and mass-to-light ratios (\citealp{GAMAmstar}; \citealp{lambdar}), star formation rates (e.g., \citealp{GAMAsfrs}, \citealp{lambdar}) and environmental measures (e.g., \citealp{groupcat}).

The primary defining subsample for this paper is the GAMA-II visual morphology catalogue, which consists of 7,556 objects in the GAMA equatorial regions with local flow-corrected redshifts in the range $0.002 < z < 0.06$, normalized redshift quality nQ$>2$ (i.e., good for science)\footnote{GAMA-derived spectroscopic redshifts are assigned a normalized quality parameter (nQ) corresponding approximately to increasing confidence levels on the measurement, with 4 representing the highest quality and confidence level measurements. Standard practice within the survey is to require at least quality level 2 for use in scientific analysis.}, and extinction corrected SDSS $r$ band Petrosian magnitude of $r < 19.8$ mag (see \citealp{vismorph} for further details).  In the GAMA visual morphology analysis, galaxies are classified into broad galaxy classes based primarily on their spheroid-/disk-dominated appearance and then their single-/multi-component nature (yielding broad classes E, S0-Sa, SB0-SBa, Sab-Scd, SBab-SBcd, Sd-Irr and LBS) via the consensus of a team of human classifiers (see Fig. 2 of \citealp{vismorph} for examples of each class). Note that the presence or absence of any tidal feature/disturbance is not a criterion of this classification scheme, so in the case of merging systems each galaxy distinguishable as a separate GAMA object is separately classified according to the aforementioned scheme. \citet{vismorph} classify 868 total galaxies as LBSs, which is the full sample of LBSs we consider here (see example LBS classification images in Fig. \ref{fig:classims}).\footnote{Note that an independent morphology classification performed by members of the SAMI Galaxy survey team also exists for a subset of these galaxies, and these independent classifications do not agree for all objects. The GAMA classifications are based on $giH$ colour images compared to $gri$ in the SAMI case, and SAMI classifications explicitly use signs of star formation as an indication of later type. These differences result in a tendency towards more later type classifications in the SAMI case (see \citealp{Rob_class} for a detailed discussion of these differences). Further, the SAMI classifications do not inlude a separate ``LBS'' class, so the LBS sample that overlaps SAMI typically falls into late-type categories in the SAMI morphology classification.}

In subsequent sections we explore the properties of LBSs in order to investigate the defining characteristics of the LBS class. We use a variety of GAMA catalogue data on the LBSs and other low mass GAMA galaxies in order to explore differences and similarities in the parameter distributions amongst them. To this end, we consider GAMA-derived stellar population parameters including colours and star formation rates, plus environments. We then extend the study to more detailed consideration of the structure and kinematics of LBSs. Specifically, improved optical imaging data is now available from the KiDS survey \citep{KiDS} and we use this to determine the detailed structural characteristics of the LBSs, in terms of both single and two-component S\'ersic fits. These fits are then compared to similar structural fits of other galaxy types. We also consider integral field spectroscopic observations for a subsample of LBS galaxies with data available in the SAMI Galaxy survey data release 2 (\citealp{SAMI_ins}; \citealp{SAMI_targ}; \citealp{SAMIdr2}) to assess whether these galaxies display primarily rotation- or dispersion-dominated kinematics.

\begin{figure}
\centering
\includegraphics[width=0.45\textwidth]{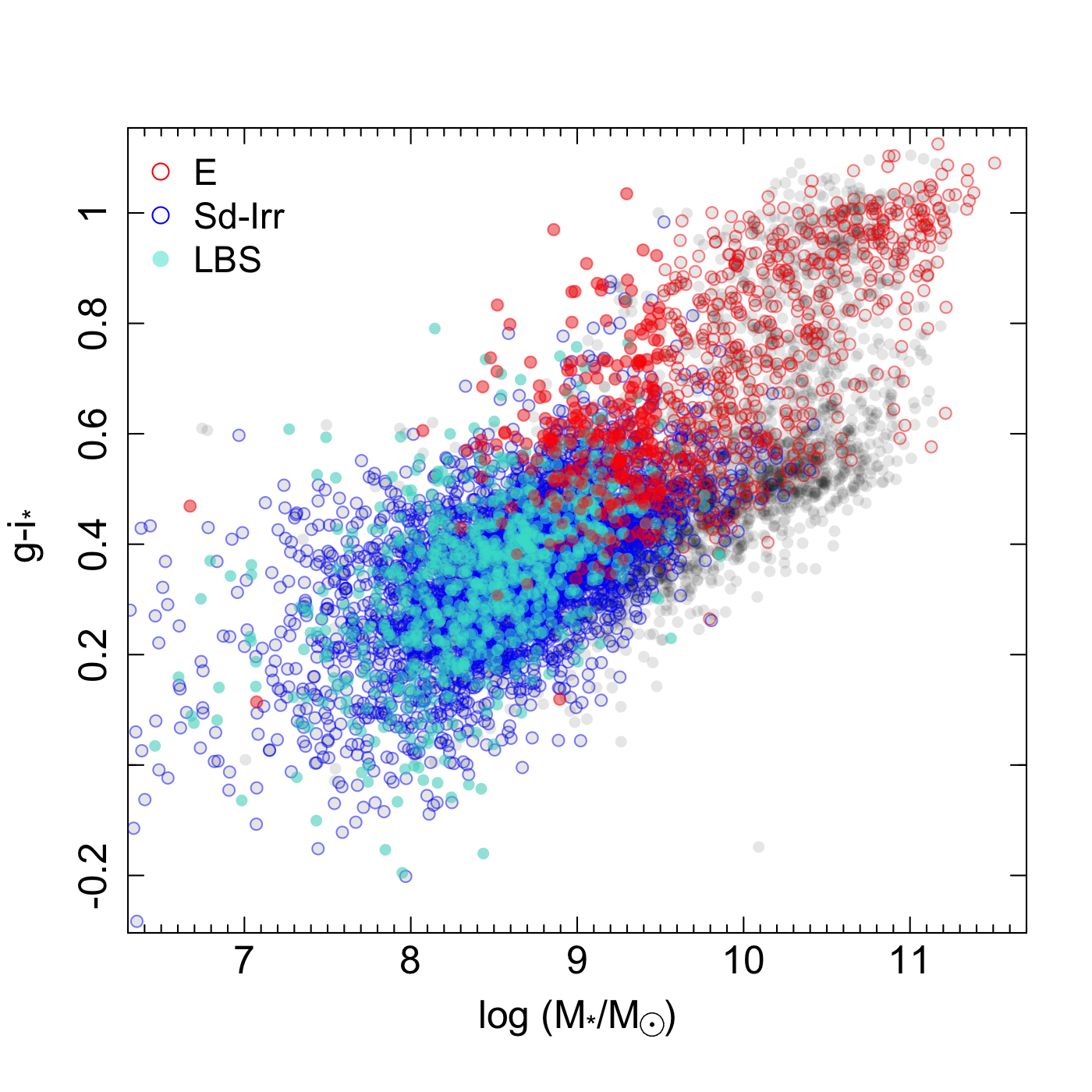}
\caption{\label{fig:colmstar} GAMA Visual Morphology sample in colour vs. stellar mass space, where $g-i$$_{*}$ is the intrinsic (corrected for internal dust extinction) $g-i$ colour from the SED modeling of \citet{GAMAmstar}. Stellar mass estimates are also derived from \citet{GAMAmstar}. Light grey points indicate the full sample distribution with coloured points indicating the E, Sd-Irr, and LBS classes. The subsample of ``low-mass E'' galaxies is indicated in filled red points.}
\end{figure}

\section{GAMA-derived Properties of LBS galaxies}
\label{prop}

We first compare the basic parameters of LBSs to those of other morphologically defined galaxy classes. In total 868 galaxies are classified as LBS in our visual morphology catalogue, and Fig. \ref{fig:colmstar} shows their position in a $g-i$ colour vs. stellar mass diagram compared to the other galaxy types. It is evident that they largely occupy the region towards the low-mass end of the blue cloud, though a few may lie on the extension of the red sequence to low mass if the spread in colour is not simply due to random error. To first order, then, it seems that the large majority of LBSs should be generically similar to faint late type galaxies (both being ``little'' and ``blue''). Further, in previous studies of GAMA dwarfs that have included members of the LBS population, such galaxies have been largely considered a star forming population inhabiting primarily low density environments (e.g., \citealp{Brough2011}; \citealp{Bauer2013}; \citealp{Mahajan2015}).

\begin{figure}
\centering
\includegraphics[width=0.45\textwidth]{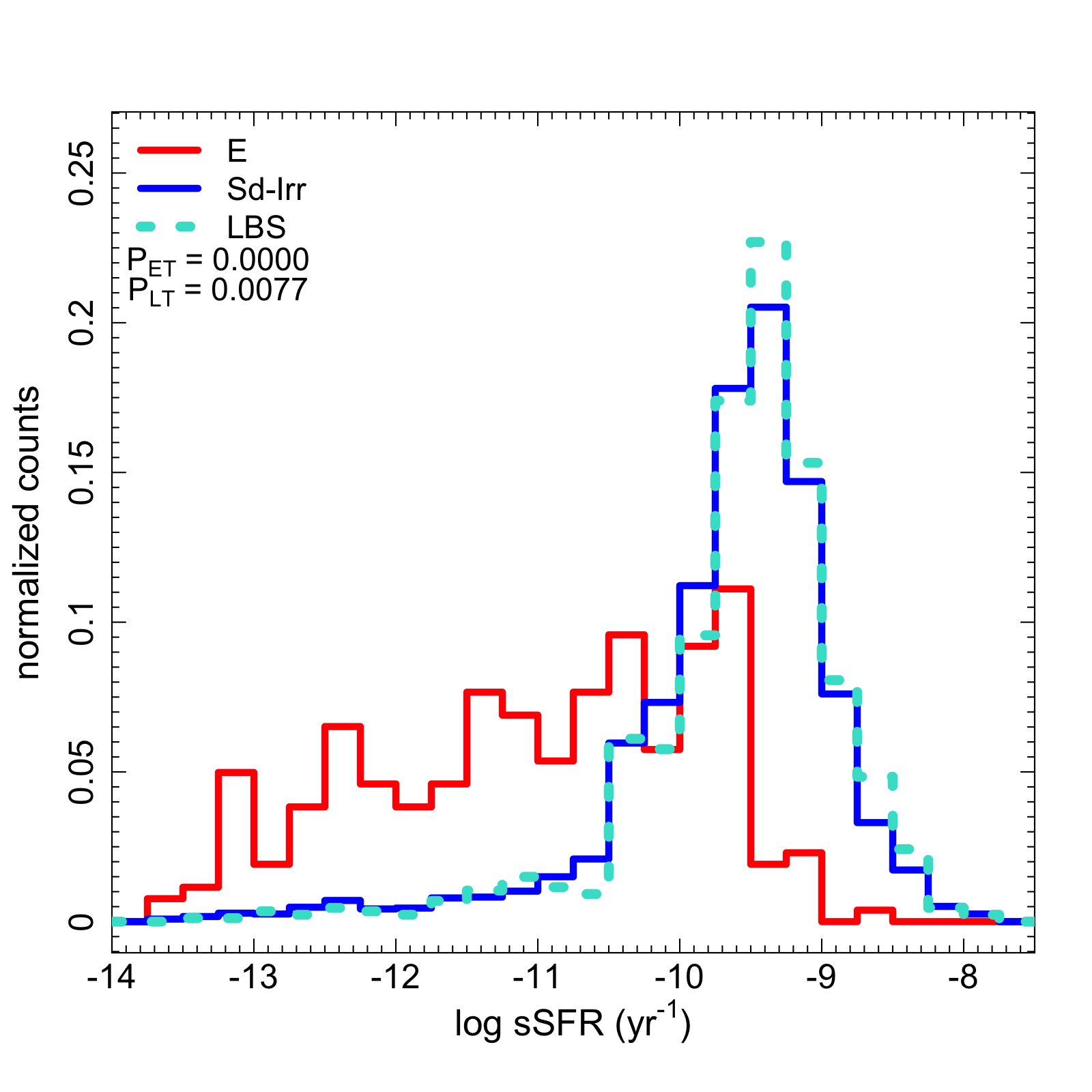}
\caption{\label{fig:LBSssfr} Distribution of LBS specific star formation rates compared to those of Sd-Irr and low-mass E galaxies, illustrating similar LBS and Sd-Irr star formation levels. The legend indicates p-values derived from K-S tests comparing the LBS property distribution to those of low-mass E (P$_{\rm ET}$) and Sd-Irr (P$_{\rm LT}$) populations.}
\end{figure}

\begin{figure}
\centering
\includegraphics[width=0.45\textwidth]{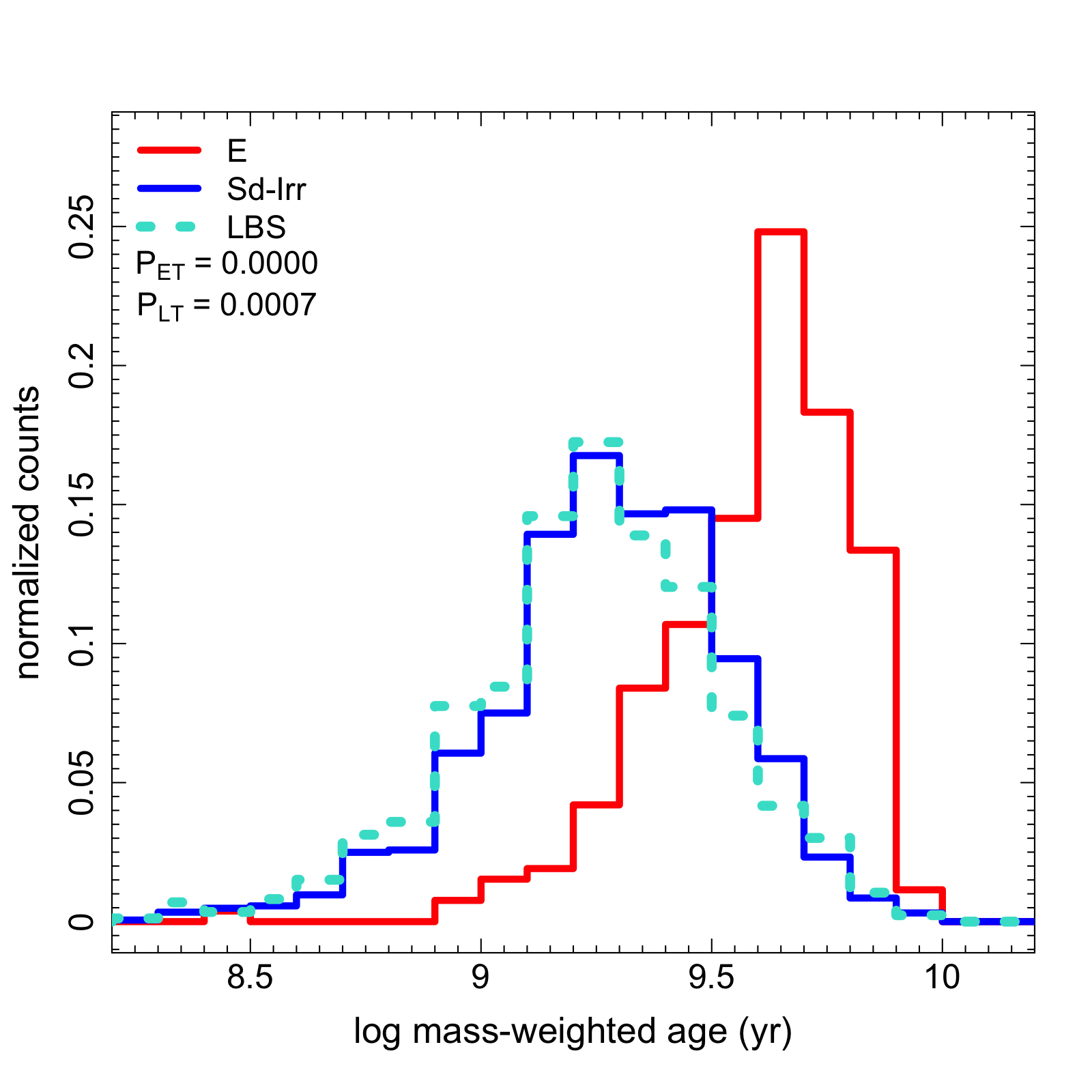}
\caption{\label{fig:LBSage} Distribution of LBS mass-weighted ages compared to those of Sd-Irr and low-mass E galaxies, illustrating a tendency towards lower ages for LBSs compared to Sd-Irr and, much more significantly, low-mass E populations. The legend indicates p-values derived from K-S tests comparing the LBS property distribution to those of low-mass E (P$_{\rm ET}$) and Sd-Irr (P$_{\rm LT}$) populations.}
\end{figure}

\begin{figure}
\centering
\includegraphics[width=0.45\textwidth]{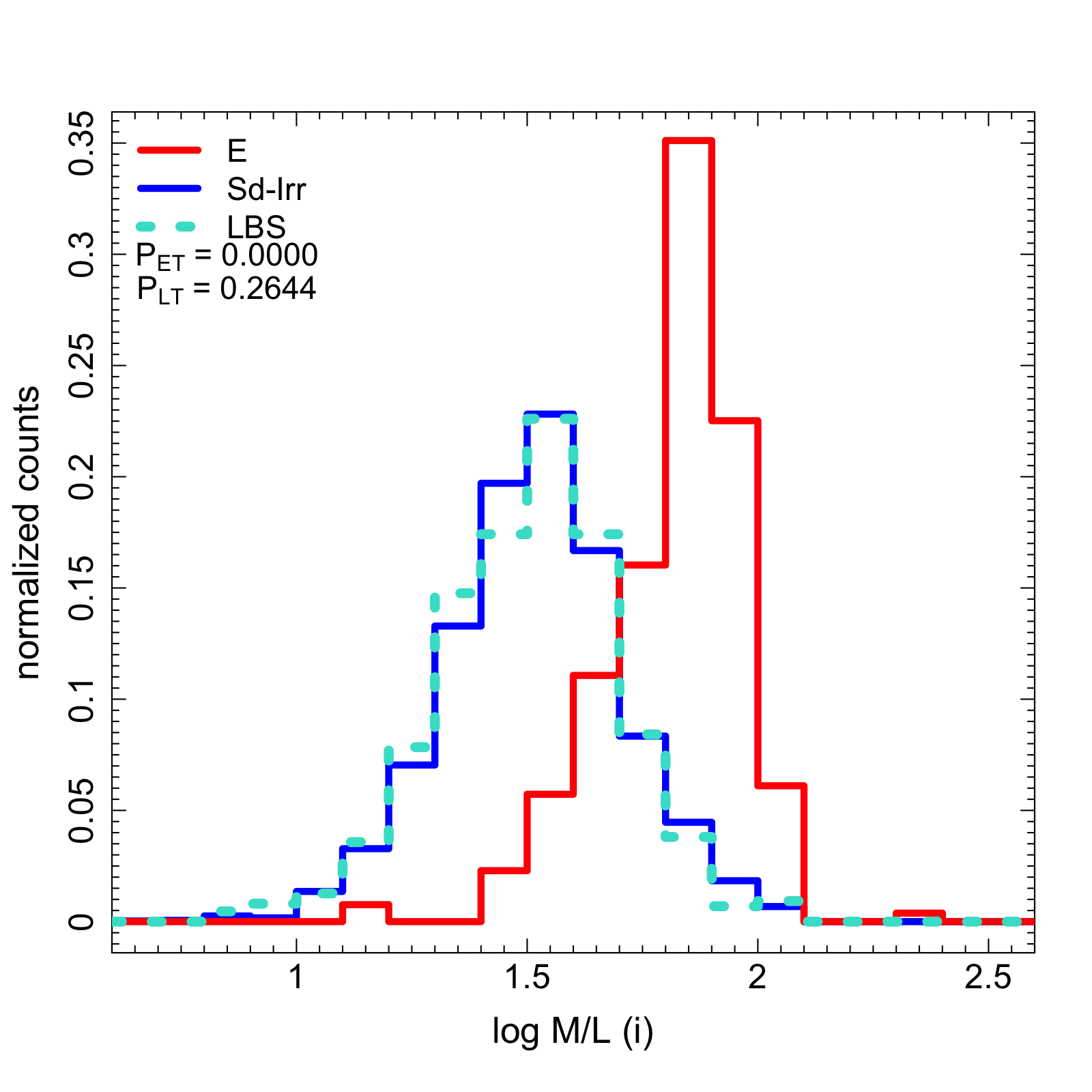}
\caption{\label{fig:LBSml} Distribution of LBS $i$-band mass-to-light ratios compared to those of Sd-Irr and low-mass E galaxies, illustrating that LBS and Sd-Irr mass-to-light ratio distributions are indistinguishable. The legend indicates p-values derived from K-S tests comparing the LBS property distribution to those of low-mass E (P$_{\rm ET}$) and Sd-Irr (P$_{\rm LT}$) populations.} 
\end{figure}

\begin{figure}
\centering
\includegraphics[width=0.45\textwidth]{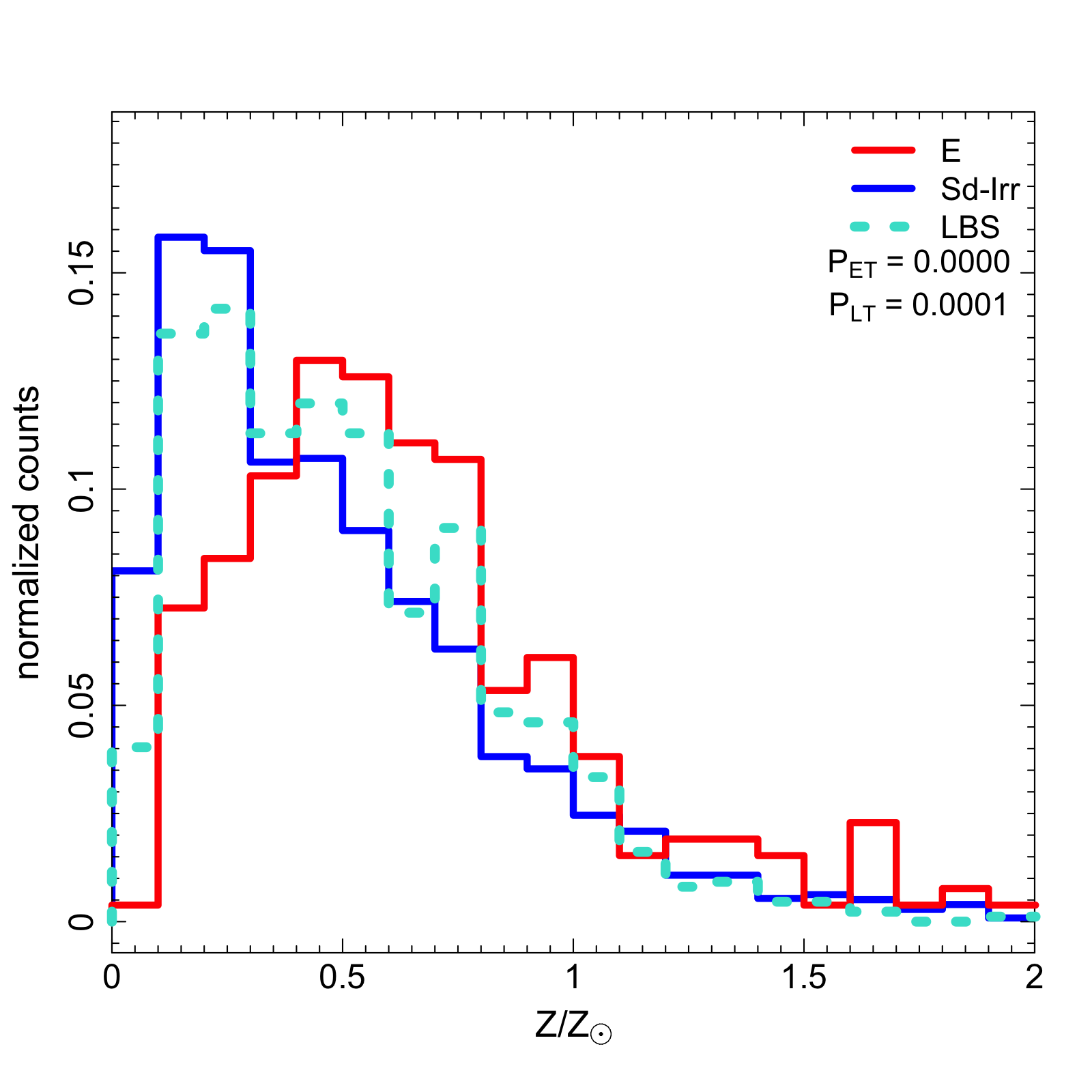}
\caption{\label{fig:LBSz} Distribution of LBS metallicities compared to those of Sd-Irr and low-mass E galaxies, illustrating that the LBS metallicity distribution skews higher than the Sd-Irr distribution but lower than the low-mass E distribution. The legend indicates p-values derived from K-S tests comparing the LBS property distribution to those of low-mass E (P$_{\rm ET}$) and Sd-Irr (P$_{\rm LT}$) populations.}
\end{figure}

However, the structure of LBSs is at odds with these areas of apparent similarity to late types. LBSs are typically compact, with median effective radius of approximately one kpc \citep{Kelvin_class}, and \citet{LangeCat} have found that the mass vs.\ size relation of LBSs was more likely compatible with Es than spirals. 

Similarly in an analysis of 73 ``blue spheroid'' galaxies, visually classified with similar criteria to LBSs as compact, spheroidal, and blue but selected from a more nearby GAMA subsample (z $<$ 0.02; $\sim$60\% overlap with our LBS sample), \citet{Mahajan2018} find that blue spheroids show structural similarity to early-type galaxies along with stellar population properties more similar to late-type galaxies. Further, \citet{Mahajan2018} find that their blue spheroid galaxies follow the same star-formation rate vs.\ atomic gas mass scaling relation as other star-forming galaxies and suggest that they could grow into spiral galaxies if supplied with sufficient gas accretion.

As illustrated in Fig. \ref{fig:colmstar}, nearly all LBSs in our sample lie at stellar masses below 10$^{9.5}$~{\rm M$_{\odot}$} (median stellar mass of the LBSs is 10$^{8.5}$~{\rm M$_{\odot}$}). While this mass distribution is similar to that of the Sd-Irr class, it is clearly different from that of the full E class. As a result, when comparing the properties of LBSs to those of Es, we choose to specifically compare to the low-mass end of the E population (also known as dwarf elliptical or dE galaxies), which we select as those below 10$^{9.5}$ in stellar mass (filled red points). Note that the LBS mass distribution does extend to lower stellar masses than the E population mass distribution. As a result, we also test a more restrictive mass selection that includes only those galaxies in the narrower mass overlap region of 10$^{8.8}$~{\rm M$_{\odot}$} $<$ M$_{*}$ $<$ 10$^{9.5}$~{\rm M$_{\odot}$}. When we redo comparisons between the three morphological classes using this selection, we find small differences in the exact values of test statistics we compute but no difference in the statistical significance of the results we report in the following sections.

In Fig. \ref{fig:LBSssfr}, we examine the GAMA-derived specific star formation rates (sSFRs) for LBSs compared to those of other visually classified late and early type galaxies. We use the stellar population fits to GAMA 21-band photometry (far-UV to far-IR) derived by \citet{lambdar} using the MAGPHYS software \citep{MAGPHYS}, where sSFRS here are averaged over 10 Myr timescales. We can see that, again, LBSs appear very similar to Sd-Irr galaxies in terms of star formation, with typical values of sSFR $\sim10^{-10}$yr$^{-1}$ (cf. \citealp{Bauer2013}), where typical sSFRs for the low-mass E population is approximately an order of magnitude lower. Though typical LBS sSFR values are much more similar to the Sd-Irrs than Es, a Kolmogorov-Smirnov (K-S) test yields a $<$1\% chance that the LBS and Sd-Irr sSFRs are drawn from the same distribution. Note that while some of the LBS sSFRs are moderately high, they are generally not as extreme as other star forming dwarfs such as BCDs, which have SFRs ranging up to a few times 10~M$_{\odot}$yr$^{-1}$ \citep{BCDsfr} compared to $<$5~M$_{\odot}$yr$^{-1}$ for LBSs, or compact star forming galaxies, with sSFRs reaching over $10^{-8}$yr$^{-1}$ \citep{CGsfr}. Thus if the high sSFR LBSs represent dwarf starbursts akin to BCDs, then the intensity of the star formation events appears to be lesser in LBSs. 

Similarly in Fig. \ref{fig:LBSage}, we find that the median mass-weighted ages of the stellar population fits from MAGPHYS are similarly distributed between LBSs and Sd-Irrs, with low-mass Es shifted to significantly higher ages. In fact, LBSs have a tendency towards slightly lower typical ages than Sd-Irrs, and the K-S test probability that these populations are drawn from the same distribution is $<$1\%. We can also see a similar effect in the $i$-band mass-to-light ratios \citep{GAMAmstar} for the three populations in Fig. \ref{fig:LBSml}. Here LBSs and Sd-Irrs appear similar but disparate from the Es, which have a much higher mass-to-light ratios due to their older (hence faded) stellar populations. K-S test probabilities reinforce that the LBS and Sd-Irr mass-to-light ratio distributions are indistinguishable, while the E distribution is formally distinct. Using the MAGPHYS results again, we examine the metallicity distributions of the three groups in Fig. \ref{fig:LBSz}. There are sizeable errors in individual metallicity estimates here, so we cannot confidently distinguish a trend in the median metallicities for each class, however we do find that the LBS and Sd-Irr metallicity distributions are unlikely to be drawn from the same distribution (K-S probability $<$1\%).

\begin{figure}
\centering
\includegraphics[width=0.45\textwidth]{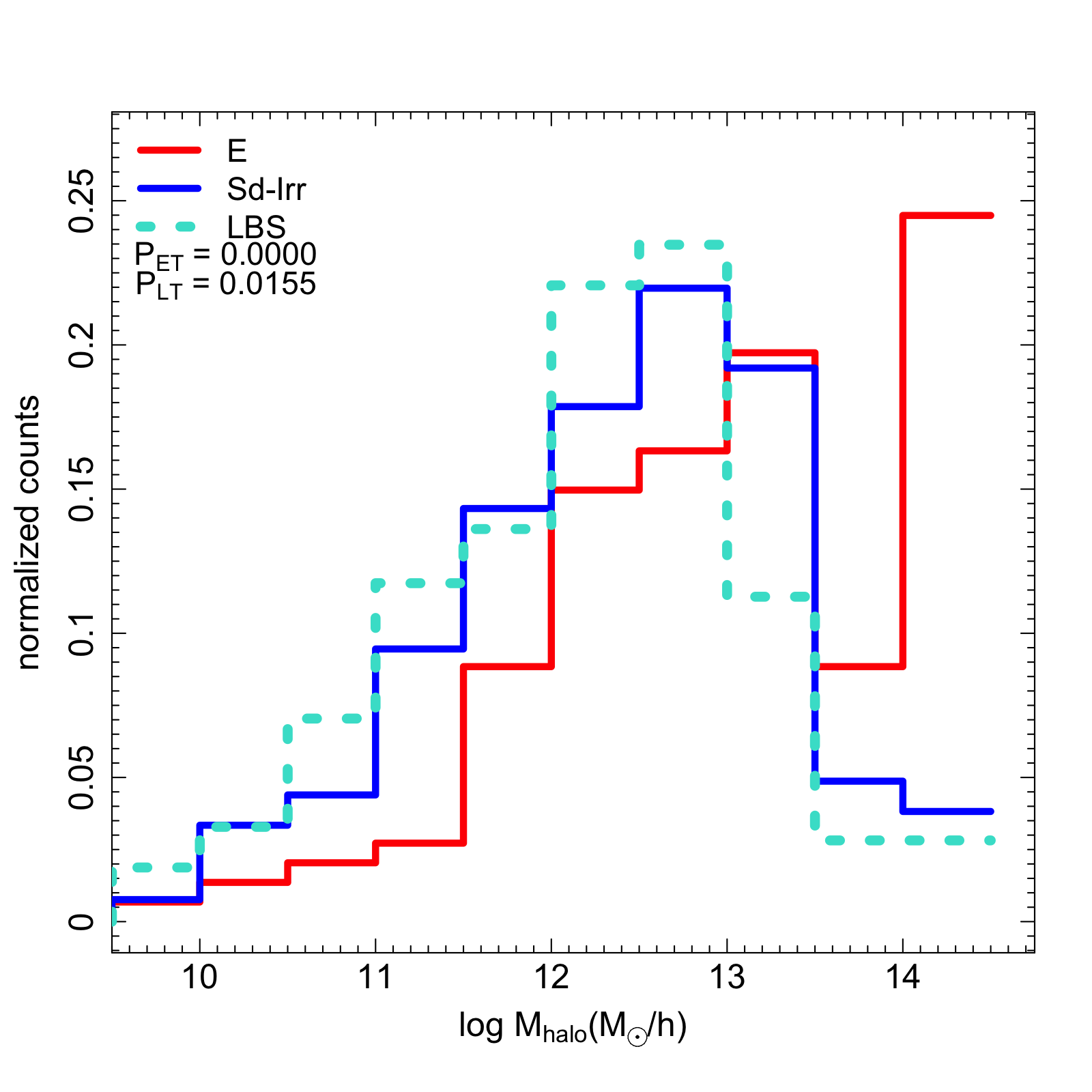}
\caption{\label{fig:LBSmhalo} Distribution of LBS group halo masses compared to those of Sd-Irr and low-mass E galaxies, where LBSs tend to inhabit lower group halo mass environments than low-mass E galaxies and potentially slightly lower group halo mass environments than Sd-Irr galaxies as well. The legend indicates p-values derived from K-S tests comparing the LBS property distribution to those of low-mass E (P$_{\rm ET}$) and Sd-Irr (P$_{\rm LT}$) populations.}
\end{figure}

Finally, we compare the local environments of LBSs to Sd-Irr and low-mass E galaxies. Fig. \ref{fig:LBSmhalo} illustrates the distribution of group halo masses derived from the GAMA survey group catalogue of \citet{groupcat}. LBS and Sd-Irr group halo mass distributions are clearly more similar than the low-mass E distribution, which is significantly shifted towards higher mass halos as is generally expected for dE galaxies (e.g., \citealp{Virgo_dEs}). A K-S test reveals an approximately 1\% chance that the LBS and Sd-Irr group halo mass distributions are drawn from the same population. Note that not all GAMA galaxies are associated with groups in this catalogue; those that do not lie in identified groups are considered ``isolated.'' Only about 35$\pm{3}$\% of the low-mass E galaxies are considered isolated by this metric, while 58$\pm{1}$\% of Sd-Irr and 65$\pm{1.5}$\% of LBS galaxies are isolated. From this GAMA group catalogue, we can also investigate the pair fractions of each galaxy class, with pairs defined by a projected physical separation of 100 kpc h$^{-1}$ or less and a velocity separation of 1,000 km s$^{-1}$ or less. We find an approximately 27$\pm{1.5}$\% pair fraction among LBSs with this metric, while Sd-Irr galaxies have a slightly higher 32$\pm{1}$\% pair fraction. Thus, in general LBSs appear to occupy slightly lower density environments than even the relatively poor environments typical of Sd-Irrs.

\section{LBS galaxy structure}
\label{struc}
We now compare the structure of LBS galaxies with Sd-Irr and low-mass E galaxies using structural parameters derived from photometric fits to VST KiDS survey (\citealp{KiDS}; \citealp{KiDSdrs}; \citealp{KiDSdr3}) $r$-band images.

\subsection{PROFIT Model Fits}

\begin{figure*}
\begin{tabular}{c}
\subfloat[]{\includegraphics[width = 6.5in]{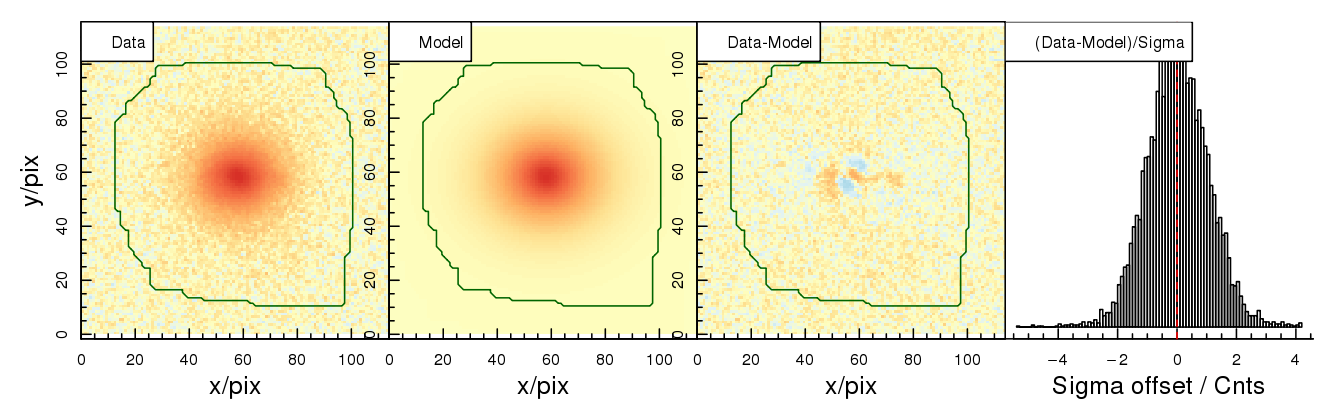}} \\
\subfloat[]{\includegraphics[width = 6.6in]{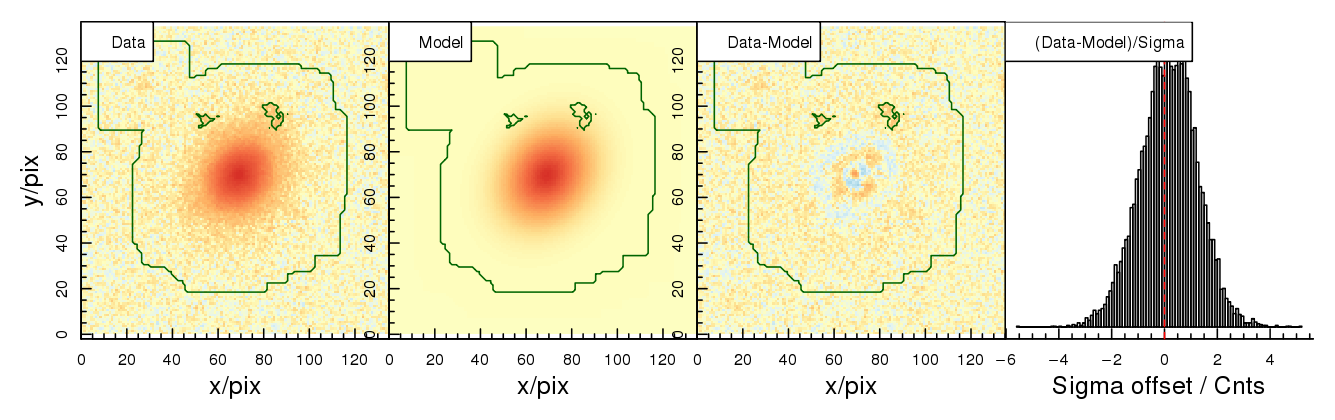}} \\
\subfloat[]{\includegraphics[width = 6.5in]{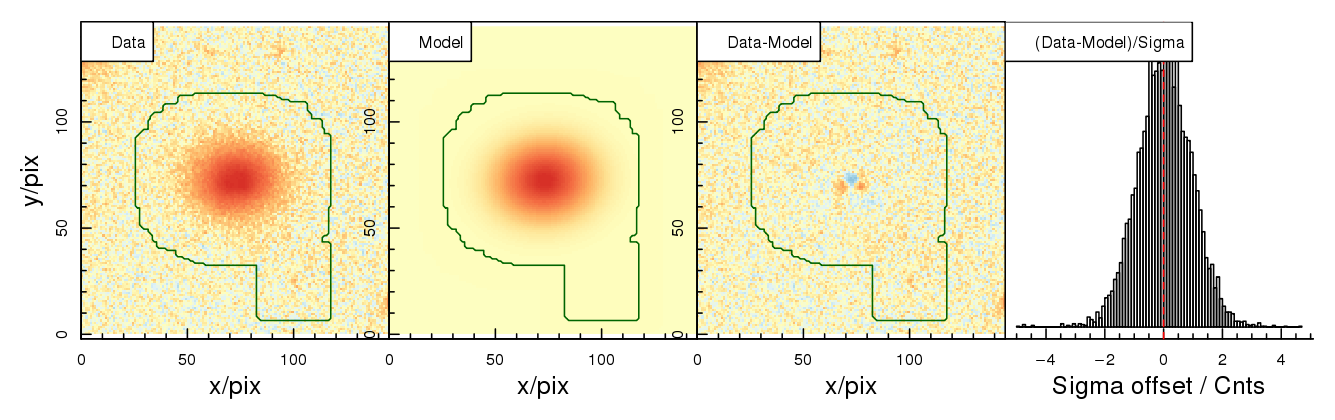}} 
\end{tabular}
\caption{Example PROFIT single S\'ersic fits to VST KiDS $r$-band images of the galaxies shown in Fig. \ref{fig:classims}. Column one shows the data image, column two shows the selected model, column three shows a difference image, and column four shows a histogram of the residuals. The green contour indicates the fitting region derived from a Source Extractor image segmentation.}
\label{fig:fits}
\end{figure*}

\begin{figure*}
\centering
\includegraphics[width=0.95\textwidth]{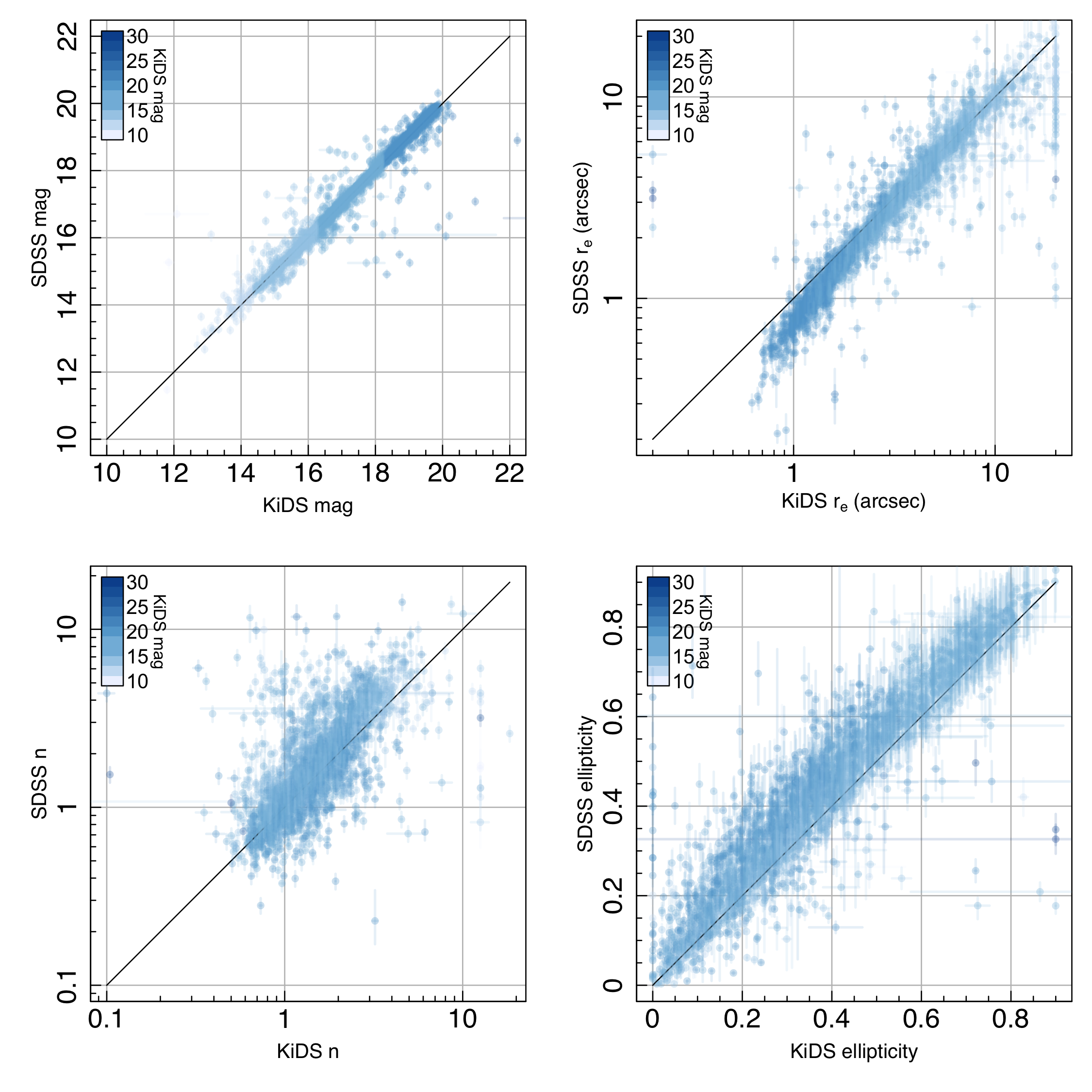}
\caption{\label{fig:single} Comparison of single S\'ersic fit parameters derived from PROFIT with VST KiDS imaging (current work) and GALFIT with SDSS imaging \citep{LangeCat}. In general, there is a close correspondence between PROFIT/KiDS and GALEX/SDSS results, although a slight tendency towards smaller measured size and lower ellipticity is seen in the PROFIT/KiDS analysis.}
\end{figure*}

We make use of the Bayesian two-dimensional profile-modeling code PROFIT \citep{PROFIT} and a series of automated wrapper codes to perform the necessary preparatory steps to running PROFIT on an input list of GAMA galaxies. PROFIT is open source (available at github.com/ICRAR/ProFit) and flexible with the ability to model a wide variety of standard model profiles and to employ an array of user-selected optimization algorithms.

To create the input images for PROFIT, we begin with calibrated, pipeline-processed $r$-band images from the VST KiDS survey data releases (\citealp{KiDSdrs}; \citealp{KiDSdr3}). We then apply additional processing steps as follows. We create an initial 400$^{\prime\prime}$ on a side cutout image (and matching VST KiDS mask image containing pipeline data quality flags) and apply a local background subtraction to the cutout image based on the LAMBDAR \citep{lambdar} background subtraction procedure. We then derive an empirical PSF from the background subtracted image using PSFEx \citep{PSFEx}. We then create a reduced-size cutout image centred on the target galaxy (sized at three times the GAMA aperture radius for each galaxy) and a matching segmentation mask of objects in the frame detected with Source Extractor \citep{sextractor}.

We next use PROFIT to obtain a single S\'ersic model fit for each object, with initial parameter guesses derived from the GAMA aperture catalogue \citep{lambdar}. We fit for the centre $x$ and $y$ pixel positions, total magnitude, r$_{e}$, S\'ersic index (n), position angle, and axial ratio parameters of each single S\'ersic model, with the r$_{e}$, n, and axial ratio parameters fit in log space. We use the centre coordinates, position angle, and axial ratio of each galaxy's GAMA catalogue aperture as our initial guess on the model's centre position, orientation, and axis ratio. Our initial guess for r$_{e}$ is one sixth of the major axis radius of the GAMA aperture, and our initial guess for the total magnitude is the SDSS DR7 catalog \citep{SDSS_dr7} Petrosian $r$-band magnitude. We also set the initial guess on the S\'ersic index equal to one. In general, these parameter guesses are somewhat arbitrary, but we structure our fitting procedure to limit sensitivity to initial guesses through use of Markov chain Monte Carlo (MCMC) sampling. We next perform an initial coarse optimization using the \textsf{R} \textsf{optim} function with the ``L-BFGS-B'' algorithm. The results of this fit are used to provide improved initial guesses to the $LaplacesDemon$\footnote{https://github.com/LaplacesDemonR/LaplacesDemon} package, which is used in MCMC mode, with the Componentwise Hit-And-Run Metropolis (CHARM) method, to find the most likely model over at least 10$^{4}$ iterations. The CHARM sampling algorithm was selected for its ability to sample across distant points in parameter space and perform well even in the presence of multimodal parameter distributions. We estimate parameter values and uncertainties only from the final stationary sample distributions, discarding up to 5000 iterations from the burnin phase, otherwise no explicit pruning of chains was performed. We also check that acceptance rates are considered suitable for our algorithm.

Finally, at the conclusion of the single S\'ersic fits for each galaxy, we use the outputs of the single S\'ersic fits to prepare initial parameter guesses for a double S\'ersic model fitting run. Here we fit for the shared centre position of both components plus bulge and disk magnitudes, radii, position angles, and axial ratios. We also fit for the bulge n parameter but fix the disk n equal to one. We use the single S\'ersic model fit centre and position angle as initial guesses for the corresponding parameter in both components. In general for our sample that is numerically dominated by low-mass galaxies, we find that the single component fit most often traces a disky galaxy component, so we use the total magnitude of the single fit as the initial guess for the disk component magnitude and half of this flux as the initial guess for the bulge component. Similarly, we use the single r$_{e}$ as the initial guess for the disk radius and half this value as the initial guess for the bulge radius. We use twice the single n value as the guess for the bulge n value. We also use an initial axial ratio value for the bulge equal to one and the single fit value as the initial guess for the axial ratio of the disk component. We then calculate the most likely double S\'ersic model according to the same fitting procedure as in the single S\'ersic case.

Using this procedure, we have performed single and double S\'ersic fits (in the KiDS $r$ band) for all of the GAMA galaxies classified as ``little blue spheroid'' from the GAMA II visual morphology catalogue \citep{vismorph} plus all GAMA galaxies that overlap with the public SAMI Galaxy survey sample target list (3159 objects in total; see example LBS fits in Fig. \ref{fig:fits}). Of these, fits for 419 galaxies initially failed because all galaxy pixels were within a masked region determined by the KiDS team (typically due to proximity to a bright star or its reflected light halo). For these initially failed fits, we found that many of these objects appear sufficiently uncontaminated that fits are possible, although some caution is necessary in interpreting the derived parameters. We subsequently fit these galaxies by providing an altered, no-masked-pixel mask image to PROFIT, and results from this analysis indicate that the majority of these objects can be reasonably well fit with this approach. However, we do assign such galaxies a quality flag indicating possibly compromised fits. For 94 objects found to have bad segmentation masks through visual inspection, an alternative solution of creating segmentation masks based directly on GAMA catalogue aperture positions has been attempted. Through visual inspection of the resulting fits, we find that some 80\%  of these objects are recoverable with reasonable fits through this approach, however again we flag all such fits as potentially compromised in quality.

As a cross check on our derived structural parameters, we compare to the prior GAMA structural fitting results of \citet{LangeCat}, based on shallower, lower-resolution SDSS imaging, in Fig. \ref{fig:single}. For overlapping objects between these two samples the single S\'ersic fit parameters are in reasonably good agreement overall, however small systematic offsets in the derived radii and ellipticity values are apparent. Since these analyses were derived from different images sources (SDSS and VST KiDS), it is plausible that these differences result from imperfections in the empirically derived PSFs between sources. However, because in the following analysis we only interpret these results comparatively within one set of data, any systematics between data sources should not affect these comparative results. Further, our results on the structure of LBS galaxies are qualitatively consistent with those derived using the earlier \citet{LangeCat} GALFIT-based structural fits.

\begin{figure}
\centering
\includegraphics[width=0.45\textwidth]{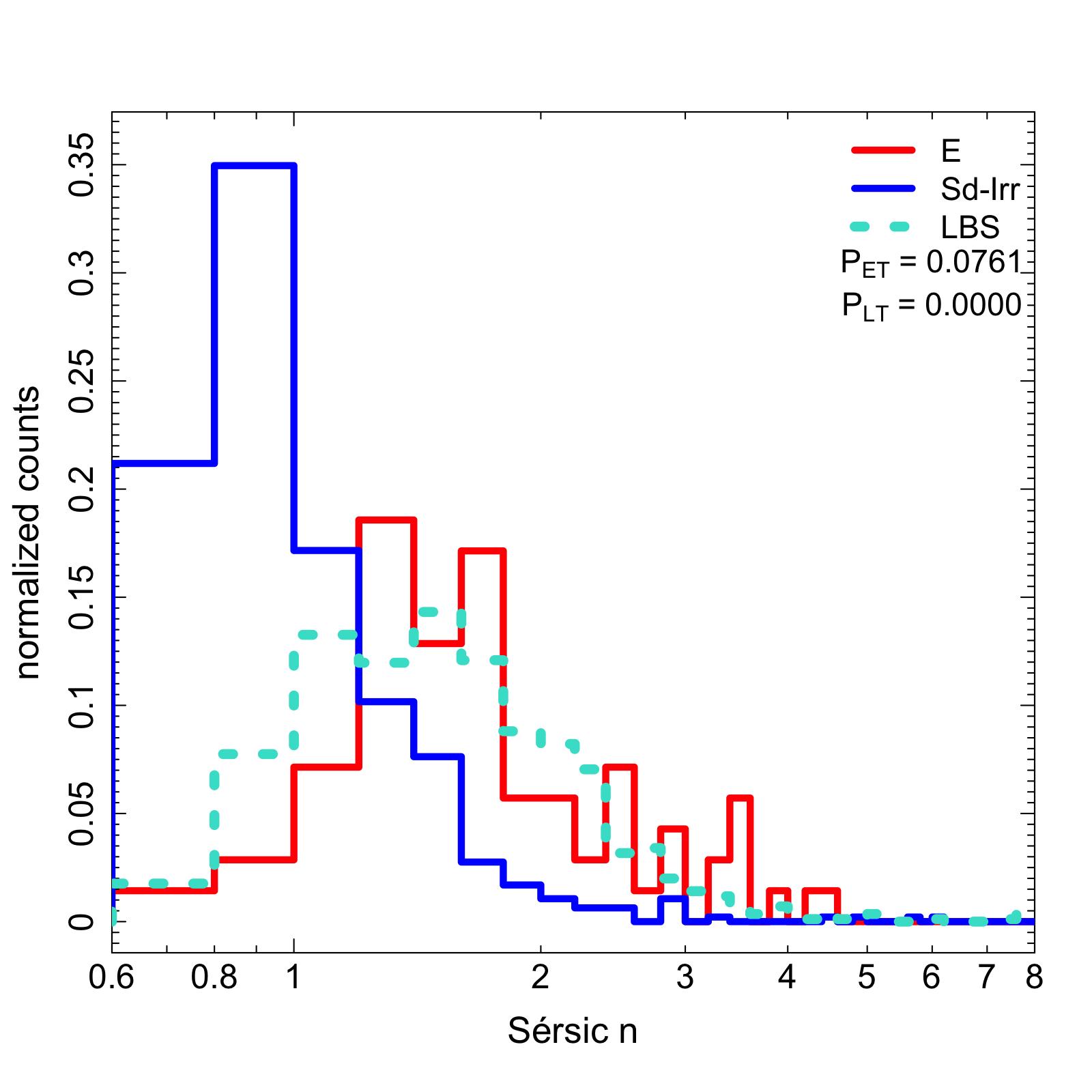}
\caption{\label{fig:LBSnhist} Distribution of LBS single S\'ersic n values compared to Sd-Irr and low-mass E galaxies, illustrating that the LBS S\'ersic n distribution is most similar to that of low-mass Es, albeit somewhat skewed to lower values. The legend indicates p-values derived from K-S tests comparing the LBS property distribution to those of low-mass E (P$_{\rm ET}$) and Sd-Irr (P$_{\rm LT}$) populations.}
\end{figure}

\begin{figure}
\centering
\includegraphics[width=0.45\textwidth]{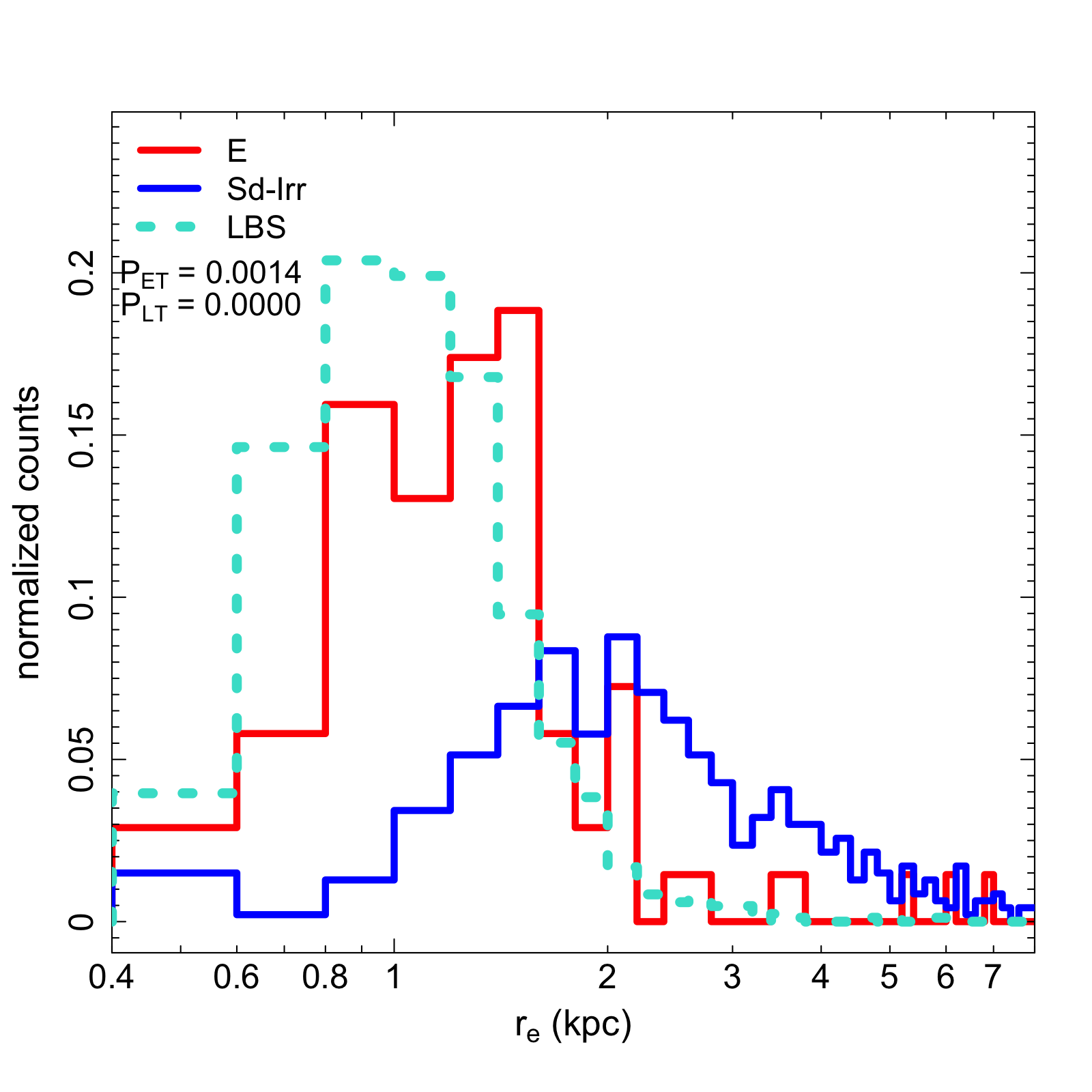}
\caption{\label{fig:LBSrehist} Distribution of LBS effective radius values (in kpc) compared to Sd-Irr and low-mass E galaxies, illustrating that the LBS effective radius distribution is similar to that of low-mass Es with a slight skew towards smaller size. The legend indicates p-values derived from K-S tests comparing the LBS property distribution to those of low-mass E (P$_{\rm ET}$) and Sd-Irr (P$_{\rm LT}$) populations.}
\end{figure}

\begin{figure}
\centering
\includegraphics[width=0.45\textwidth]{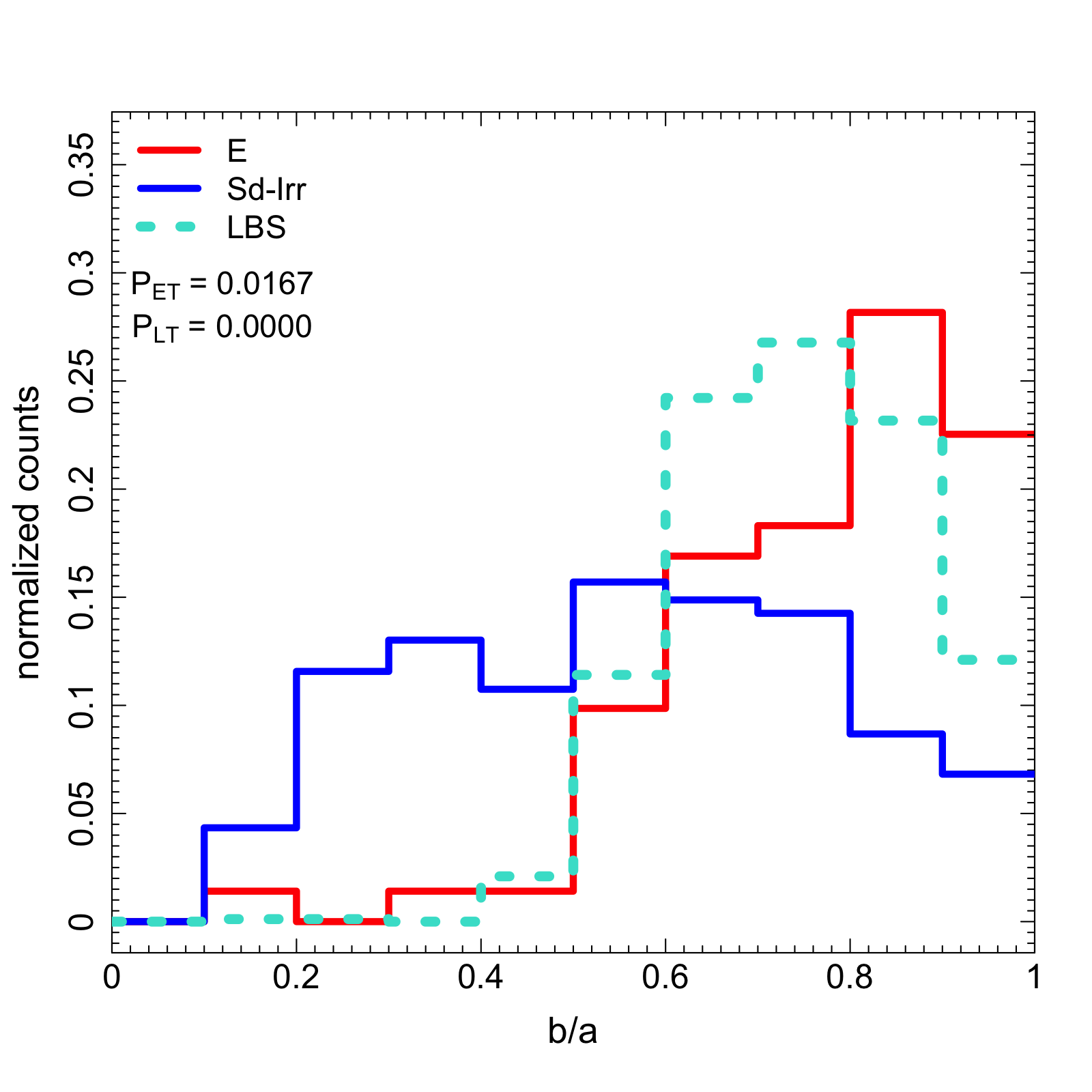}
\caption{\label{fig:badist} Distribution of LBS axial ratios compared to Sd-Irr and low-mass E galaxies, illustrating that the LBS axial ratio distribution is more similar to low-mass Es than Sd-Irrs. The legend indicates p-values derived from K-S tests comparing the LBS property distribution to those of low-mass E (P$_{\rm ET}$) and Sd-Irr (P$_{\rm LT}$) populations.}
\end{figure}

\subsection{Structural Model Fit Results}
 
We now use these two-dimensional model fits to investigate the detailed structure of the LBS galaxies. Note that although galaxies are complex and, in many cases, multi-component structures, in light of the extra parameter degeneracies inherent in multi-component modelling and the difficulty of deciding how many components are required for a satisfactory fit in an automated manner, we choose to focus here on our most stably measured single-component equivalent parameter values as a basis for comparison between the populations under consideration. Fig. \ref{fig:LBSnhist} illustrates the $r$-band single S\'ersic n distributions of LBS, Sd-Irr, and low-mass E galaxies. We find that LBSs have a similar S\'ersic index distribution to that of low-mass Es. Although the LBS distribution appears to skew slightly lower, the median S\'ersic n value for these two classes is consistent within uncertainties (1.63$\pm$0.6 vs.\ 1.69$\pm$0.6 for LBSs and Es, respectively).  On the other hand, the LBS n distribution skews higher than that of Sd-Irr galaxies, which have a median n of 0.94$\pm$0.3 (i.e., steeper radial profiles), and K-S test results also emphasize that these distributions differ significantly. However, comparing with a common early-/late-type divider used for giant galaxies, approximately 86\% of LBSs have S\'ersic n $<$ 2.5, which would typically be associated with disks or late types, and the other $\sim$14\% have n $>$2.5, which would typically be associated with bulges or early types. We check that the S\'ersic values for the low-mass E sample matches the expectation for classical dE galaxies from the Virgo Cluster. It is well known that n decreases with decreasing luminosity (e.g., \citealp{YC94}; \citealp{GG03}), and most Virgo dEs in our luminosity/mass range have n between 0.5 and 3 (e.g., \citealp{Gavazzi05}), consistent with typical values for both our low-mass Es and LBSs.

Fig. \ref{fig:LBSrehist} illustrates the size distributions of three classes, where the LBS population skews to smaller r$_{e}$ than either Sd-Irrs or low-mass Es (K-S probability $<$1\% for LBS compatibility with either distribution). However, LBSs have a median r$_{e}$ of $\sim$1 kpc, which is consistent with the median r$_{e}$ of low-mass Es within uncertainties. We can also see in Fig. \ref{fig:badist} that consistent with the morphological impression of spheroidal shape, the axial ratio (b/a) distribution of LBSs appears more similar to that of low-mass Es than to the Sd-Irrs (K-S probability $<$1\% for LBS compatibility with Sd-Irr distribution but $\sim$1\% for low-mass E distribution). We see that Sd-Irr galaxies have the wide spread of axial ratios, with median $\simeq 0.5$, expected of a disk population (cf. \citealp{SFS70}; \citealp{AR02}), while the spread of LBS axial ratios appears lower. The median b/a for LBSs is also consistent with that of low-mass Es within uncertainties (0.74$\pm{0.13}$ and 0.80$\pm{0.17}$ for LBSs and Es, respectively).

Thus we confirm the incongruous results that LBSs have stellar population properties similar to Sd-Irr galaxies but structural properties that are compatible with low-mass Es. As the ellipticity measures only 2-D shapes, one plausible option at this point might be that, rather than being a true spheroidal population, LBSs are just near face-on late types, perhaps with only a low level of irregularity and a more centrally peaked profile than usual, which are unneccessarily selected out as a separate class. To test this idea, we check that the Sd-Irr class does not seem to be lacking round images compared to the Sab-Scd class and find that adding the Sd-Irr and LBS classes would give an overall mean ellipticity lower than that of Sab-Scds (i.e., early-type spirals would be less round than late-type spirals, which seems unlikely). We have also checked that Sd-Irr galaxies with higher values of n have the same ellipticity distribution as those with lower n (and hence still different to the LBSs). 

Since LBS galaxies were originally morphologically classified from SDSS data, our deeper VST KiDS imaging offers the opportunity to identify any lower surface brightness features that may have been poorly detected at SDSS depth. We do find some examples of LBSs with more obvious two-component structure in this new imaging dataset, and to quantify this, we consider a likelihood ratio test with the null hypothesis that the single S\'ersic model is a good description for each galaxy. We then calculate the probability of this null hypothesis according to a chi-square distribution with degrees of freedom given by the difference in degrees of freedom between the two models. If this probability is $<$5\%, we consider the double S\'ersic model to be preferable. We find that this metric implies $\sim$85\% of LBSs are better represented by two-component models. By this same metric, we find that only $\sim$30\% of our low-mass E sample would be best described by two-component models (previous observations of dwarf early types with embedded disk components include the works of e.g., \citealp{Jerjen00}; \citealp{Barazza02}; \citealp{DeRijcke03}; \citealp{Geha03}; \citealp{GG03}). Thus, the more common presence of disk components in LBS galaxies may be responsible for the small differences in structural parameters compared to low-mass E galaxies, and we investigate the possibility of disk-like structure in LBSs further through consideration of their kinematics in the next section.

\section{SAMI Kinematics of LBSs}
\label{SAMIcomp}

We now examine the SAMI Galaxy survey integral field kinematics for LBS galaxies with data available in SAMI Data Release 2 \citep{SAMIdr2}. The SAMI survey selection is primarily drawn from GAMA, and there are 62 LBS galaxies that have emission line data of sufficient quality for spatially resolved kinematic analysis available in SAMI data release 2. For comparison, we also consider a sample of elliptical galaxies (20) and Sd-Irr galaxies (60) from our GAMA morphology sample with similar masses to the LBS sample and high-quality emission line data available from SAMI. These emission line data are processed using the LZIFU \citep{LZIFU} line fitting procedure described by \citet{SAMIdr1} and \citet{Medling18}. We use the velocity maps derived from this analysis to extract rotation curves along each galaxy's major axis position angle, which we determine using the kinemetry methods of \citet{kinpa}. Note that each SAMI velocity field covers a footprint $\sim$15$"$ in diameter, which for the median size of our LBS galaxies covers out to $\sim$5r$_{e}$. We fit the derived one-dimensional rotation curves using the following simple piecewise functional form for velocity as a function of radius (e.g., \citealp{Wright07}; \citealp{Epinat09}):

\begin{equation}
V(r)=
\begin{dcases}
  V_{t}\times(r/r_{t})            & r \leq r_{t}\\
  V_{t}           & r > r_{t} \text{.}
\end{dcases}
\end{equation}

For comparison, we also use SAMI stellar kinematics measurements \citep{SAMI_stellkin}, specifically the stellar velocity dispersion values measured within elliptical r$_e$ derived using the Multi-Gaussian Expansion method (MGE; \citealp{MGE1}; \citealp{MGE2}) and code from \citet{Scott09} applied to GAMA imaging (d'Eugenio, priv. comm.). We note that comparing rotational velocity derived from ionized gas with stellar velocity dispersion is not necessarily standard practice since stars and gas may not trace the underlying potential in the same way, e.g., due to asymmetric drift. Here the decision to do so is largely a practical one; since the low-mass LBS galaxies are star forming, the emission line data provide superior S/N to trace the kinematics to large radii. \citet{Cortese14} find the average relationship V$_{\rm rot}$(stars)$/$V$_{\rm rot}$(gas) is 0.75 for SAMI galaxies, and we find that a correction to our measured gas rotation velocities of this magnitude would not substantially change the following results.

\begin{figure*}
\centering
\includegraphics[width=0.98\textwidth]{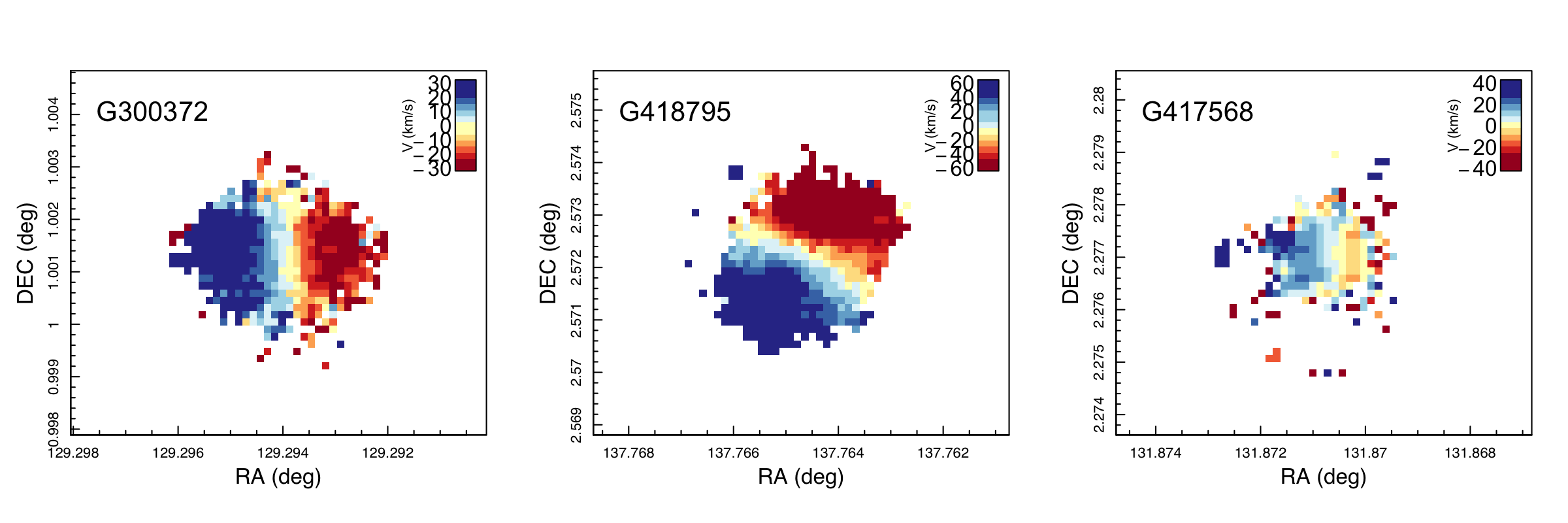}
\caption{\label{fig:LBSvex} SAMI emission line velocity maps derived from the analysis described by \citet{SAMIdr1} and \citet{Medling18}. The left and middle panels show examples of LBS galaxies with velocity fields that display extended, regular rotation well fit by the symmetric rising rotation curve form used here, while the far right panel shows an example of a LBS velocity field that is poorly fit with this rotation curve form. Note that the category of galaxies that are poorly fit by a symmetric rising rotation curve (and for which we cannot obtain a reliable characteristic rotation velocity with this method) includes both objects with some evidence of disturbed rotation (as in the far right panel above) and those with no apparent rotation signature. We omit spaxels with large measured velocity errors ($>$15 km$/$s) from the plotted maps.}
\end{figure*}

\begin{figure*}
\begin{tabular}{cc}
\subfloat[]{\includegraphics[width = 3.5in]{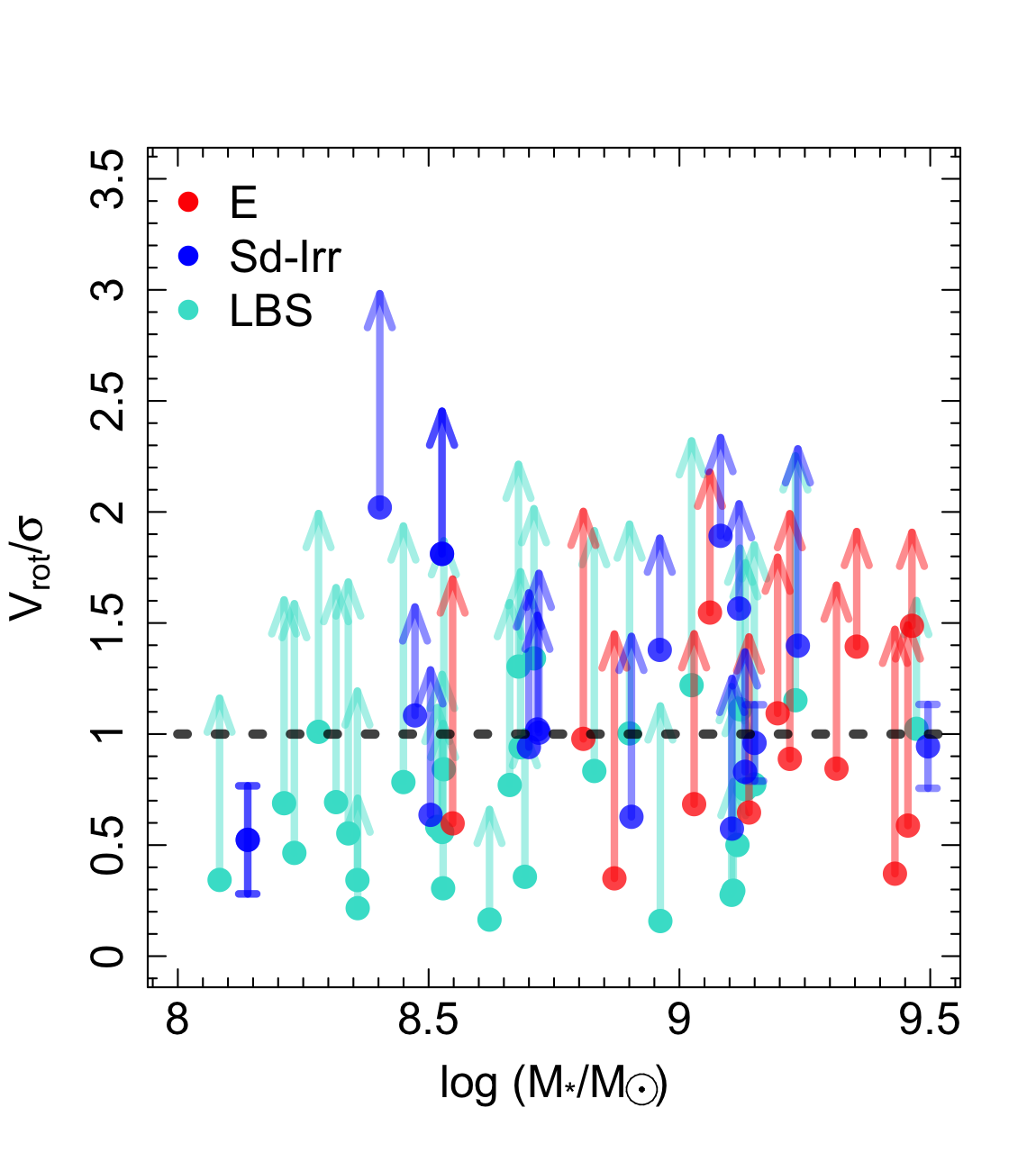}} &
\subfloat[]{\includegraphics[width = 3.5in]{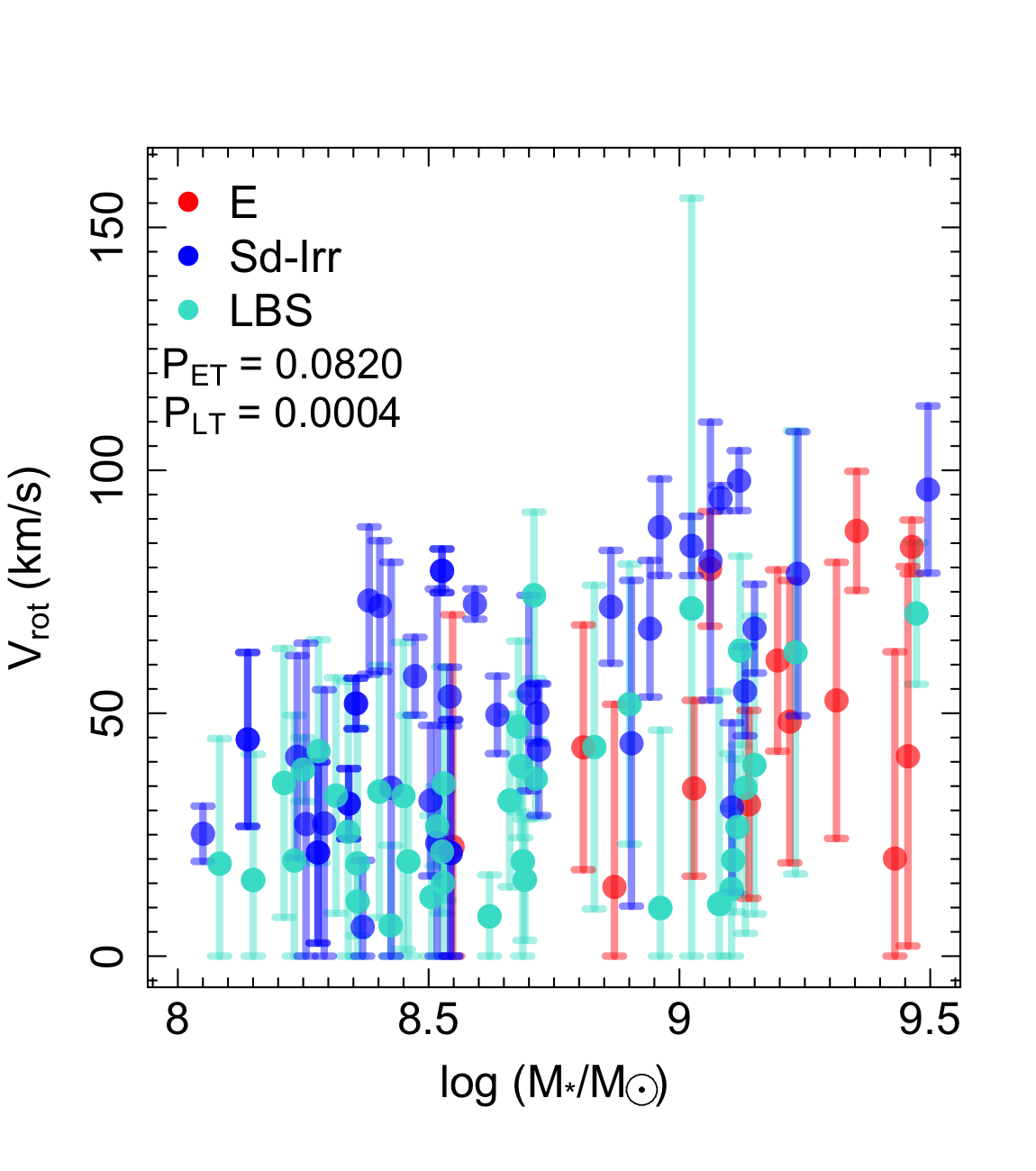}}
\end{tabular}
\caption{\label{fig:LBSvsigma} Panel a: SAMI-derived rotational velocity divided by velocity dispersion (V$_{\rm rot}$$/$$\sigma$) versus stellar mass for LBS and low-mass E and Sd-Irr galaxies. Error bars are plotted as limit arrows where the SAMI stellar velocity resolution is greater than the measured velocity dispersion. The black dashed line indicates equal contribution from rotational velocity and velocity dispersion. LBSs and low-mass Es for which we can derive reliable rotation measures are typically consistent with or above this line, and as such we infer they are at least marginally rotation-dominated systems. Panel b: SAMI-derived, rotational velocities versus stellar mass for LBS and low-mass E and Sd-Irr galaxies. At constant stellar mass, Sd-Irr rotation velocities appear to be slightly higher than LBS and E velocities, which are more similar. The legend indicates p-values derived from K-S tests comparing LBS rotation velocities to those of the low-mass E (P$_{\rm ET}$) and Sd-Irr (P$_{\rm LT}$) samples.}
\end{figure*}

Out of the 62 LBS galaxies with emission line kinematics maps, we can derive an ionized gas maximum circular velocity (V$_{\rm rot}$) for 42 galaxies with this method. The remaining 20 LBSs have either ill-defined rotation axes or disturbed kinematics that are not well fit by the simple rising rotation curve form used here (see Fig. \ref{fig:LBSvex} for rotating and disturbed velocity field examples). Fig. \ref{fig:LBSvsigma}a shows the derived ionized gas rotation velocity divided by stellar velocity dispersion (V$_{\rm rot}$/$\sigma$) versus galaxy stellar mass for the LBSs and low-mass Es and Sd-Irrs with measurable rotation curves and measurable stellar velocity dispersion values as determined by \citet{SAMIdr2}, totalling 31 LBSs, 13 low-mass Es, and 19 low-mass Sd-Irrs. Rotation velocities are inclination corrected based on estimating inclination from each galaxy's ellipticity and assuming intrinsic flattening of 0.3. Since the stellar velocity resolution of SAMI data is $\sim$70 km~s$^{-1}$, the $\sigma$ values we derive for these low-mass galaxies typically represent upper limits, and we represent these cases with arrows replacing their error bars in the V$_{\rm rot}$/$\sigma$ space. We find that both LBSs and low-mass Es with measurable rotation typically have V$_{\rm rot}$/$\sigma$ consistent with or greater than unity. In fact, the V$_{\rm rot}$/$\sigma$ distributions of all three morphological classes appear similar at these low masses, with the prevalence of limit values on $\sigma$ in this dataset making it problematic to determine whether or not any class has a significant offset in typical V$_{\rm rot}$/$\sigma$.

  Although we find that some of our LBSs have disturbed or not clearly rotating kinematics, the V$_{\rm rot}$/$\sigma$ values we derive for the others indicate that $\sim$68\% of our LBS galaxies are at least marginally rotation-dominated systems. We find that a similar fraction of our low-mass Es are also rotation dominated. Significant rotation has previously been observed in the \emph{stellar} kinematics of early-type galaxies and, in particular, appears to be more common for low-mass early types (see e.g., \citealp{Emsellem07}; \citealp{Brough17}; \citealp{VDS2017A}; \citealp{Rob_class}). It is also not surprising that Sd-Irr galaxies in this sample are typically found to be rotation dominated.

In Fig. \ref{fig:LBSvsigma}b, we further explore the characteristic rotation velocities for members of each class with measurable rotation. We see that at constant stellar mass, the Sd-Irr galaxies appear to have slightly larger V$_{\rm rot}$ values than LBS galaxies, with a $<$1\% K-S test probability that these populations are drawn from the same distribution. In contrast, LBS V$_{\rm rot}$ values appear to be more similar to those of low-mass E galaxies, however we cannot conclusively say these V$_{\rm rot}$ distributions are identical, as we find $\sim$8\% K-S test probability that these populations are drawn from the same distribution.

\section{Discussion and Conclusions}
\label{conc}

To summarize the results of the preceding sections, we find that the dwarf ``Little Blue Spheroid'' (LBS) galaxy class combines stellar population properties similar (although not identical) to spiral/irregular galaxies (including colours, specific star formation rates, stellar population ages, mass-to-light ratios, and metallicities) with morphology and structural properties (including S\'ersic indices, radii, and axial ratios) compatible with low-mass elliptical galaxies. Further, LBSs typically occupy relatively poor environments similar to Sd-Irr galaxies but with an even greater tendency towards isolation. From analysis of SAMI kinematics of LBSs, we also find that the \emph{majority} of LBSs display at least marginally rotation-dominated dynamics, similar to low-mass ellipticals.

We first consider whether or not this population is a plausible star-forming, field-environment precursor to dwarf elliptical galaxies. We then discuss the overlap of this population with other compact, star-forming populations and the evolutionary processes that are potentially associated with such populations, including interactions/mergers, external gas accretion, and downsizing in galaxy formation.

Our PROFIT modeling analysis suggests that LBSs are structurally equivalent to dwarf elliptical galaxies but with ongoing star formation. In recent work, \citet{George17} has also found structural similarity between dwarf ellipticals and a 55-galaxy sample of SDSS star-forming, blue early-type galaxies. The star formation (history) of LBSs seems to be very similar to that in the equally low mass but disk-like faint Sd-Irr galaxies (which we can equate to dI galaxies). Adding in the fact that they live in similar low density environments as dIs, we can then hypothesize that LBSs are field counterparts of cluster dEs. They have continued to form stars at a relatively constant rate (given that the inverse of their sSFR is of order 10$^{10}$ years) while their clustered cousins have largely ended their star formation much earlier. In turn this dichotomy suggests that, just as at high masses, there are two structural types of low-mass galaxy, spheroidal and disk-like, but that star formation continues in both types absent a quenching mechanism. Given their similar sSFRs to dwarf irregular galaxies, it seems likely that the LBSs have sufficient fuel to maintain substantial star formation for several more Gyr. Indeed, though we do not as yet have gas masses for our LBS sample, \citet{Mahajan2018} do find that their blue spheroids (including many of our lowest redshift LBSs) display similar gas properties to star-forming disk galaxies.

If these LBSs were to decrease or end their star formation, we would expect them to move towards the region of the colour vs. stellar mass diagrams occupied by low-mass Es (see Fig. \ref{fig:colmstar}). This would suggest that the progenitors of cluster dEs could be LBS-like objects and not dwarf irregulars, which could solve the long standing problem of how to transform star forming dwarf irregulars to create the large cluster dE populations, given the differences in structure and surface density between the two classes (e.g., \citealp{DP88}; \citealp{Meyer14}). As we have found here, LBSs already show strong structural similarity to low-mass Es, so quenching of star formation brought on by the effects of infall into a richer environment could be sufficient to make LBSs similar to dEs, whereas a structurally different dI would have to be both quenched and have its structure altered by environmental interaction. We find that most LBSs have significant rotational support and some low-mass Es also display similar kinematic properties, implying minimal change to the dynamical structure of LBSs would be required to transition into such low-mass Es. However, due to their previously discussed low density environments, it is unlikely that the currently observed LBSs will directly transform into their clustered dE cousins. Interestingly, \citet{Janz2017} has identified examples of isolated and quenched low-mass early-type galaxies with rotating kinematics, which suggests early-type dwarfs can also be quenched outside of rich environments.

Blue compact dwarf (BCD) galaxies have frequently been suggested as a progenitor population to dEs (e.g., \citealp{DP88}; \citealp{Meyer14}), and there is significant but not perfect overlap between LBSs and the typical colour, surface brightness, and magnitude range used to select BCDs (e.g., \citealp{BCDdef}). Morphologically there is also significant similarity between these galaxy types, with compact cores frequently surrounded by a lower surface brightness component. Like BCDs, star formation in our LBSs also appears primarily centrally concentrated, although extended disk star formation could remain undetected at the depth of our SAMI maps. LBSs also overlap with the luminous blue compact galaxy and Green Pea classifications in their compact, star-forming nature (e.g., \citealp{LCBG}; \citealp{peas}), however the typical LBS is less massive, less extremely star forming, and found at lower redshift than these populations.

Interactions between galaxies are a common explanation for the mixture of compact morphology and recent star formation observed in galaxies like BCDs (e.g., \citealp{TT95}; \citealp{Pust01}; \citealp{Noeske01}). Similarly, the blue E/S0 population of \citet{KGB} is strongly associated with interactions or mergers. More recently compact dwarf populations have been suggested as the likely products of \emph{dwarf-dwarf} mergers that could both drive a central starburst and lead to disturbed kinematics in the remnant galaxy (e.g., \citealp{Lelli12}; \citealp{Ashley14}; \citealp{Koleva14}). However, as previously noted, the typical star formation rates of LBSs are lower than those typical of BCDs, and only $\sim$10\% of our LBSs reach sSFRs compatible with starbursts. Within the stellar mass range most of our LBSs inhabit these galaxies make up $\sim$20\% of the galaxy population \citep{vismorph}, and therefore the high sSFR tail represents only a few percent of the low-mass population. Interestingly, this figure is similar to the $\sim$4\% major merger rate estimated at such low masses by \citet{Casteels14}, which seems to support the plausibility of a merger origin for such galaxies. Most other LBSs would then be expected to originate from events that would incite lower intensity star formation, possibly still including interactions or more minor merger events.

External gas accretion may also be an important ingredient in maintaining the star formation we observe in LBSs. \citet{Graham17} consider in detail a case of an isolated dwarf early-type galaxy that closely overlaps the properties of our LBS class. In this galaxy, the authors find not only significant rotation but also gas and stellar components that are counter-rotating with respect to one another, strongly implying an external accretion origin for the galaxy's gas supply. External accretion that enables building of a new disk around an existing compact core appears to be a plausible process at work in LBS galaxies, particularly the $\sim$85\% of LBSs we find are best fit with an additional (typically disk-like) structural component.

Finally, LBSs could also potentially represent a downsized version of a galaxy population both predicted and observed at higher redshifts, the ``blue nuggets'' (e.g., \citealp{Barro2013}; \citealp{DB2014}; \citealp{Zolotov2015}; \citealp{Tacchella2016}). The blue nugget population is believed to form through a process of gas compaction triggered by mergers or significant gas accretion. The products of this process are forming stars at a rate similar to the high end of the LBS distribution (sSFR $\sim$10$^{-9}$ yr$^{-1}$) but are typically an order of magnitude higher in stellar mass with accordingly higher stellar mass surface density. Thus if created through similar processes, LBSs would have to represent a downsized (more recently formed and thus lower mass) version of this population.

In summary, we find that LBSs could plausibly emerge from a mixture of merger and accretion processes acting on the low-redshift dwarf galaxy population. These scenarios are not necessarily mutually exclusive, and our observed LBS population may well be a mixture of recent merger products, galaxies in currently isolated environments forming stars at a slow and steady pace, and galaxies with significant gas accretion building up a disk component. We also conclude that LBS galaxies resemble a non-quenched, field-environment counterpart to dEs, and if a LBS galaxy were to be subjected to larger-scale environmental forces such as infall into a richer cluster environment, it would likely evolve into a product resembling a typical cluster dE. The kinematic diversity in LBS galaxies, from disturbed to regularly rotating, also gives us a clue to the likely spread of evolutionary processes affecting this population. In future work, we plan to use the growing survey samples of integral field kinematic data for galaxies extending into the low mass regime in order to better understand the link between the photometrically inferred structure and detailed kinematic characteristics of such galaxies.

\section*{Acknowledgements}

AJM acknowledges funding support from a Vanderbilt University Stevenson Postdoctoral Fellowship and thanks Andreas Berlind and Ferah Munshi for helpful conversations. SP thanks his project students Alex Tidd and Jamie Ward for their help in the early stages of some of this work. We also thank the anonymous referee for his/her comments, which have greatly improved the clarity of this work. LC is the recipient of an Australian Research Council Future Fellowship (FT180100066) funded by the Australian Government. Parts of this research were conducted by the Australian Research Council Centre of Excellence for All Sky Astrophysics in 3 Dimensions (ASTRO 3D), through project number CE170100013. SB acknowledges the funding support from the Australian Research Council through a Future Fellowship (FT140101166). MSO acknowledges the funding support from the Australian Research Council through a Future Fellowship (FT140100255). Support for AMM is provided by NASA through Hubble Fellowship grant \#HST-HF2-51377 awarded by the Space Telescope Science Institute, which is operated by the Association of Universities for Research in Astronomy, Inc., for NASA, under contract NAS5-26555. JvdS is funded under Bland-Hawthorn's ARC Laureate Fellowship (FL140100278).

GAMA is a joint European-Australasian project based around a spectroscopic campaign using the Anglo-Australian Telescope. The GAMA input catalogue is based on data taken from the Sloan Digital Sky Survey and the UKIRT Infrared Deep Sky Survey. Complementary imaging of the GAMA regions is being obtained by a number of independent survey programmes including GALEX MIS, VST KiDS, VISTA VIKING, WISE, Herschel-ATLAS, GMRT and ASKAP providing UV to radio coverage. GAMA is funded by the STFC (UK), the ARC (Australia), the AAO, and the participating institutions. The GAMA website is http://www.gama-survey.org/ .

The SAMI Galaxy Survey is based on observations made at the Anglo-Australian Telescope. The Sydney-AAO Multi-object Integral field spectrograph (SAMI) was developed jointly by the University of Sydney and the Australian Astronomical Observatory. The SAMI input catalogue is based on data taken from the Sloan Digital Sky Survey, the GAMA Survey and the VST ATLAS Survey. The SAMI Galaxy Survey is supported by the Australian Research Council Centre of Excellence for All Sky Astrophysics in 3 Dimensions (ASTRO 3D), through project number CE170100013, the Australian Research Council Centre of Excellence for All-sky Astrophysics (CAASTRO), through project number CE110001020, and other participating institutions. The SAMI Galaxy Survey website is http://sami-survey.org/.

Based on data products from observations made with ESO Telescopes at the La Silla Paranal Observatory under programme IDs 177.A-3016, 177.A-3017 and 177.A-3018, and on data products produced by Target/OmegaCEN, INAF-OACN, INAF-OAPD and the KiDS production team, on behalf of the KiDS consortium. OmegaCEN and the KiDS production team acknowledge support by NOVA and NWO-M grants. Members of INAF-OAPD and INAF-OACN also acknowledge the support from the Department of Physics \& Astronomy of the University of Padova, and of the Department of Physics of Univ. Federico II (Naples).

\bibliographystyle{mnras}
\bibliography{LBS}

\label{lastpage}
\end{document}